\documentclass[aps,amsmath,amssymb,twocolumn,showpacs]{revtex4-1}

\usepackage{graphicx}
\usepackage{dcolumn}
\usepackage{bm}
\usepackage{color}
\usepackage{lipsum}

\begin{document}

\title{Optical properties of coupled silicon nanowires and unusual mechanical inductions}

\author{R. M. Abraham Ekeroth$^{1}$}
\email{mabraham@exa.unicen.edu.ar}

\affiliation{$^{1}$ Grupo de Plasmas Densos, Instituto de F\'{\i}sica Arroyo Seco, Universidad Nacional del Centro de la Provincia de Buenos Aires, Pinto 399, 7000 Tandil, Argentina}

\date{\today}

\begin{abstract} 
A recent study of the photonic coupling between metallic nanowires has revealed new degrees of freedom in the system. Unexpected spin torques were induced on dimers when illuminated with linearly polarized plane-waves. As near-field observables, the spectra of torques showed more resolved resonances than typical far-field spectra. Here the study is extended to silicon dimers. Strong forces and torques are exerted by light under both polarizations $s$ and $p$, contrary to plasmonic systems where the resonant strong forces are found only for $p$-polarization. The systems made of high-dielectric possess volume resonances that induce the forces differently than in plasmonic systems, which have surface resonances. The asymmetry in strong near-fields is responsible for the unusual mechanics of the system. Some consequences of that may include the breaking of the action-reaction principle or the appearance of pulling forces. 

The numerical study is based on an exact method. The work is thought for the design of nanorotators and nanodetectors. It suggests a new viewpoint about optical forces: the resultant dynamics of topological variations of electromagnetic fields.
\end{abstract}

\maketitle  

\section{Introduction}

Light is known to exert forces and torques on mesoscale objects \cite{Novotny}. In general, the transfer of both linear and angular momenta to the object is possible if complex beams of light are used \cite{AndrewsBookStrLight}. In particular, a single plane wave having linear polarization exerts only radiation pressure on a single object that pushes it into the forward direction. Furthermore, this force can be found to have resonances that can be followed from Mie expansions as dependent on the geometry and constitution of the object in question \cite{Bohren1998}. However, when having two or more optically coupled objects, the interaction between them makes the realistic scattering very complex in general \cite{Dholakia2010,Raziman2015,Nieto2015,Tsuei1988}. There are no satisfactory theoretical descriptions of all the involved phenomena.

As a growing research, the field of nanophotonics demands the knowledge of the exact consequences of light-matter interactions at nano and mesoscales. The importance of this knowledge lies, for example, in the wide possibilities already demonstrated to move, trap, or guide subwavelength objects \cite{Grzegorczyk2006,Marago2013,Spesyvtseva2016,Gao2017}. Thus, the correct description of the optical forces is essential for the design of photonic-based small devices \cite{Ma2012,Merklein2017}, especially in biology \cite{Ashkin1987,Svoboda1994,Righini2009,Gongora2016}, optical matter \cite{Burns1990,Figliozzi2017,Nan2018}, optical circuits \cite{Li2008,Renaut2013} among other subfields. 

In particular, the dimer's electromagnetic scattering is very known under the so-called small particle approximation or Rayleigh regime \cite{Bohren1998}. Under this regime, the objects' response is represented by the coupling of dipole moments induced by the incident and the scattered light \cite{Dholakia2010,NunoMODimer,NordlanderPHDim}. As a result of the coupling, binding forces appear in addition to the scattering forces or radiation pressures exerted by light \cite{Lamothe2007,Albaladejo2009,Dholakia2010,Miljkovic2010}. On the other hand, the exact forces induced in coupled objects can be calculated numerically by a proper integration of the Maxwell Stress Tensor \cite{Jackson}. In particular, unexpected optical torques have been found to be exerted on metallic dimers of infinite nanowires under illumination with a single plane wave having linear polarization \cite{Abraham2016,Ding2017,Abraham2018Ag}. Surface plasmon resonances were found to induce these torques in addition to the usual components of the optical forces. The results had no precedents in the literature and they cannot be approached by small particle approximations.

In this paper, the study of the optically-induced mechanics is extended to dielectric nanowire dimers, in particular using silicon that is a very useful high-contrast dielectric \cite{NanotechHandbook}. A high-contrast dielectric can sustain electromagnetic modes that correspond to morphological dependent resonances (MDRs) \cite{VanBladel1977,Ng2005,Abraham2013}. These modes correspond to volume resonances whose optical properties are well different from surface resonances \cite{Decker2016}. For example, the former resonances have strong field concentrations inside the wires' volume while the latter ones enhance the fields around the surface of the objects. Another difference between high-dielectric and plasmonic systems is that the former systems have both strong electric and magnetic resonances while the plasmonic systems have only electric spots in general \cite{Etxarri2011,Schmidt2012,Albella2013,Bakker2015,Chen2018}. In particular, this fact corresponds to have strong resonances under both fundamental polarizations, $s$ and $p$, in two dimensional (2D) problems with high-contrast materials \cite{Mirzaei2015}.  The dielectric properties of silicon have been summarized in the Ref.~\cite{Abraham2013}, for instance. An interband direct transition induces a great enhancement of the relative dielectric function with respect to the typical value $\varepsilon_{r}\approx12.25$ that is taken for the quasistatic limit \cite{DeSousa2016}. Then, silicon objects produce strongly confined fields \cite{Caldarola2015,Vergaz2016} due to the excitation of Mie resonances and narrow electromagnetic resonances in the far fields at optical regime \cite{Xiao2010}. As no exact satisfactory model exists for systems of coupled dielectric wires \cite{Cao2011}, a complete numerical study is performed here of the MDRs in coupled wires of silicon. The aim of the work is to show how new degrees of freedom appear for the dielectric dimer when the realistic interaction between the wires is taken into account. Hybridized MDRs appear in the dimers' responses with respect to the MDRs of isolated-like objects \cite{Zywietz2015,Li2017}, as it occurs for plasmonic dimers \cite{NordlanderPHDim,Abraham2016,Abraham2018Ag}. A new relation between mechanical and field observables is realized and it shows how strong asymmetries of the near-field distribution induces strong force components or torques, or other interesting effects like the breaking of the action-reaction law \cite{Sukhov2015} or the eventual presence of pulling forces.

Although an exact electromagnetic method is used here, neither thermal nor Brownian forces are considered \cite{Albaladejo2011}. Also, no ``dynamic'' forces are calculated, i.e. forces that take into account initial velocities and accelerations of the wires \cite{Grzegorczyk2006dynamics}. Then, no complete dynamics is obtained for the system. Yet, the results are believed to approach the movement of coupled particles with more accuracy than in previous works. The spectra of the mechanical inductions can be used to design nanorotators, filters of nanosystems or nanodetectors \cite{Ramos2013}.

\section{Methodology \label{sec:Methods}}

The whole methodology used in this work was presented in previous papers \cite{Abraham2015_2,Abraham2016,Abraham2018Ag}. However, as we are now interested in morphological excitations, we must revise the methodology for the two fundamental polarizations existing in 2D scattering problems.

\subsection*{Integral Method to obtain the exact fields}

\begin{figure}[!h]
 \begin{centering}
 \includegraphics[width=9cm,height=9cm,keepaspectratio]{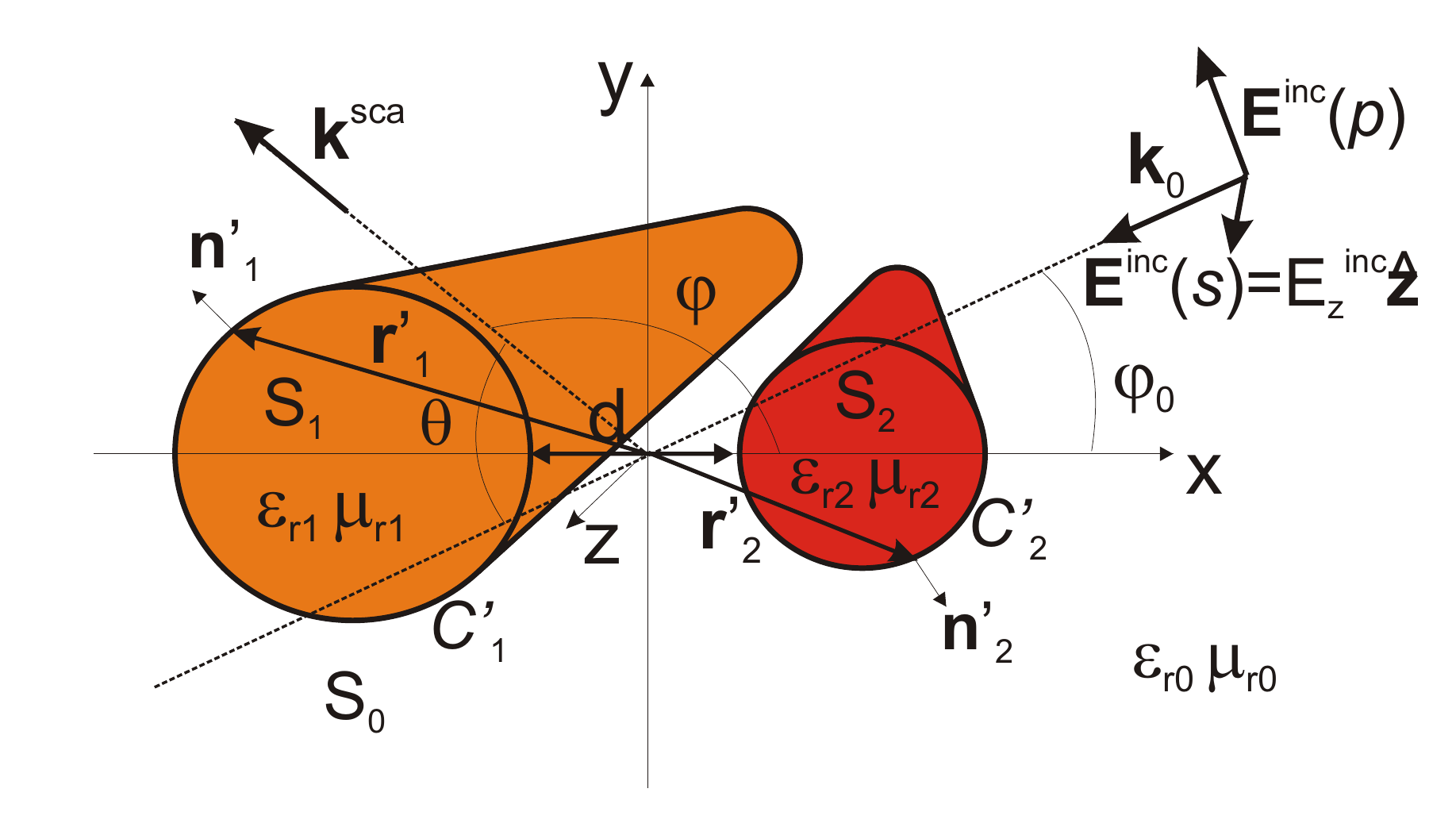}
 \caption{\label{fig:1_EsqIntMethod} Scheme of the scattering region formed by two interacting cylinders, which compose a dimer of nanowires having circular cross-sections. The contours of the wires are $C'_1$ and $C'_2$, left and right cylinder respectively. The definition of coordinate system is shown, together with the geometrical and constitutive parameters.}
 \par\end{centering}
\end{figure}

A well-known integral method is the general exact formulation used here to calculate both the near and far fields of the dimer's scattering \cite{OFTMetInt3D,MetIntCylPlane,Abraham2015_2,Lester-nanoantennas2}. Time-harmonic dependence of fields $ exp[-i\omega t]$ is assumed.  Let's suppose two coupled scatterers having axial symmetry along $z$-axis. Under this setting, two fundamental polarizations exist which can be used to express any solution of the scattering by the system. They correspond to an electric (magnetic) field aligned with the $z$-axis, namely polarization $s$ ($p$) respectively. Then, the solution for each component of the electromagnetic field is split into two families. Each family depends on a unique scalar function $\psi_{\alpha}(\mathbf{r})$ where $\alpha=s;p$ is the corresponding polarization. This function is the union of three piecewise solutions $\psi^{(j)}_{\alpha}$ on each domain $j=0,1,2$ of the dimer, respectively (see Fig.~\ref{fig:1_EsqIntMethod}). They obey the equations

\begin{align}
\left[\bigtriangledown_{t}^{2}+k_{j}^{2}\right]\psi^{(j)}_{\alpha}(\mathbf{r})&=0, \label{eq:HelmGralSimAxial} \\
\psi^{(j)}_{\alpha}(\mathbf{r})\mid_{\mathbf{r}\rightarrow C'^{(-)}_{m}}&=\psi^{(k)}_{\alpha}(\mathbf{r})\mid_{\mathbf{r}\rightarrow C'^{(+)}_{m}}  \label{eq:cc1probSimAxial} \\
\frac{1}{\nu_{rj}(\alpha)}\frac{\partial\psi^{(j)}_{\alpha}(\mathbf{r})}{\partial\hat{\mathbf{n}}}\mid_{\mathbf{r}\rightarrow C'^{(-)}_{m}}&=\frac{1}{\nu_{rk}(\alpha)}\frac{\partial\psi^{(k)}_{\alpha}(\mathbf{r})}{\partial\hat{\mathbf{n}}}\mid_{\mathbf{r}\rightarrow C'^{(+)}_{m}}, \label{eq:cc2probSimAxial}
\end{align} 
where $j$, $k$ represent two continuous and adyacent media with respect to the contour $C'_{m}$ of the scatterer $m=1,2$. Here, $\bigtriangledown_{t}$ is the transversal nabla operator and the vectors $\mathbf{r},\mathbf{k}_{0}$ also belong to the 2D space $(x,y)$. The wavevector $\mathbf{k}_{0}$ with magnitude $k_0=\frac{2\pi}{\lambda}\sqrt{\varepsilon_{r0}\mu_{r0}}$ -see Fig~(\ref{fig:1_EsqIntMethod})- is provided by the incident wavelength $\lambda$, the surrounding medium characterized by $\varepsilon_{r0}$, $\mu_{r0}$ and the incident angle $\varphi_0$. $\nu_{rj}(\alpha)$ is a polarization-dependent factor defined as $\nu_{rj}(s)= \mu_{rj}$ and $\nu_{rj}(p)= \varepsilon_{rj}$. Thus, the fields in every point in space can be obtained by the ``principal'' field $\psi^{(j)}_{\alpha}$ and its complemmentary field $\frac{i}{\omega\nu_{rj}(\alpha)\nu_{0}(\alpha)}\nabla_{t}\times \psi^{(j)}_{\alpha}(\mathbf{r})\hat{\mathbf{z}}$, where $\nu_{0}(\alpha)$ is another polarization-dependent factor defined as $\nu_0(s)= \mu_0$ and $\nu_0(p)= \varepsilon_0$. For the sake of clarity, they are specified below for each polarization, namely

\begin{align}
\mathbf{E}^{(j)}(\mathbf{r})=\psi^{(j)}_{s}(\mathbf{r})\hat{\mathbf{z}}, \label{eq:Evector_s} \\
\mathbf{H}^{(j)}(\mathbf{r})=\frac{i}{\omega\mu_{rj}\mu_{0}}\nabla_{t}\times \psi^{(j)}_{s}(\mathbf{r})\hat{\mathbf{z}}.\label{eq:HtAxial_s}
\end{align}
for $s$-polarization, and 
\begin{align}
\mathbf{H}^{(j)}(\mathbf{r})=\psi^{(j)}_{p}(\mathbf{r})\hat{\mathbf{z}}, \label{eq:Hvector_p} \\
\mathbf{E}^{(j)}(\mathbf{r})=\frac{i}{\omega\varepsilon_{rj}\varepsilon_{0}}\nabla_{t}\times \psi^{(j)}_{p}(\mathbf{r})\hat{\mathbf{z}}.\label{eq:EtAxial_p}
\end{align}
for $p$-polarization respectively. 

With the convention of normals as shown in Fig.~\ref{fig:1_EsqIntMethod}, the expressions for the scattered field on each medium $j = 0,1,2$ can be written as

\begin{widetext}
\begin{align}
\psi_{\alpha}^{(0)}(\textbf{r})=\psi_{\alpha}^{inc}(\textbf{r}) 
+\frac{1}{4\pi}\int_{C'_{1}}ds'\left[\frac{\partial G_{0}(\mathbf{r},\mathbf{r}')}{\partial\mathbf{n}'_{1}}\psi_{\alpha}^{(0)}(\textbf{r}')-G_{0}(\mathbf{r},\mathbf{r}')\frac{\partial\psi_{\alpha}^{(0)}(\textbf{r}')}{\partial\mathbf{n}'_{1}}\right] 
+\frac{1}{4\pi}\int_{C'_{2}}ds'\left[\frac{\partial G_{0}(\mathbf{r},\mathbf{r}')}{\partial\mathbf{n}'_{2}}\psi_{\alpha}^{(0)}(\textbf{r}')-G_{0}(\mathbf{r},\mathbf{r}')\frac{\partial\psi_{\alpha}^{(0)}(\textbf{r}')}{\partial\mathbf{n}'_{2}}\right], \notag \\ 
\textbf{r}\in S_{0}; \label{eq:psi0DC}
\end{align}

\begin{align}
\psi_{\alpha}^{(1)}(\textbf{r})= 
-\frac{1}{4\pi}\int_{C'_{1}}ds'\left[\frac{\partial G_{1}(\mathbf{r},\mathbf{r}')}{\partial\mathbf{n}'_{1}}\psi_{\alpha}^{(1)}(\textbf{r}')-G_{1}(\mathbf{r},\mathbf{r}')\frac{\partial\psi_{\alpha}^{(1)}(\textbf{r}')}{\partial\mathbf{n}'_{1}}\right], \textbf{r}\in S_{1}; \label{eq:psi1DC}
\end{align}

\begin{align}
\psi_{\alpha}^{(2)}(\textbf{r})= 
-\frac{1}{4\pi}\int_{C'_{2}}ds'\left[\frac{\partial G_{2}(\mathbf{r},\mathbf{r}')}{\partial\mathbf{n}'_{2}}\psi_{\alpha}^{(2)}(\textbf{r}')-G_{2}(\mathbf{r},\mathbf{r}')\frac{\partial\psi_{\alpha}^{(2)}(\textbf{r}')}{\partial\mathbf{n}'_{2}}\right],\textbf{r}\in S_{2}; \label{eq:psi2DC}
\end{align}
\end{widetext}
where $\psi^{(j)}_{\alpha}(\textbf{r})$ represent the complex amplitudes in the host media ($j=0$), or in the scatterers' volume, i.e. $j=1,2$ respectively. If the module of the electric incident field $\left|\psi_{s}^{inc}\right|$ is assumed to be known, for instance, then the scalar complex function $\psi_{s}^{inc}(\mathbf{r})=\left|\psi_{s}^{inc}\right|e^{-i\mathbf{k}_{0}.\mathbf{r}}$ represents the incident field under $s$-polarization. The incident electric field under $p$-polarization can be obtained by the plane-wave relationship of the fields to obtain $\psi_{p}^{inc}(\mathbf{r})$. The integral equations (\ref{eq:psi0DC}-\ref{eq:psi2DC}) are solutions of Eq.~(\ref{eq:HelmGralSimAxial}). $ds'$ denotes differential length element over $C'_{1}$ or $C'_{2}$. The $G_{j}$'s are the Green functions, which are solutions of the inhomogeneous Helmholtz equation

\begin{equation}
\nabla^{2}G_{j}(\mathbf{r},\mathbf{r}')+k_{0}^{2}\epsilon_{rj}G_{j}(\mathbf{r},\mathbf{r}')=-4\pi\delta(\mathbf{r},\mathbf{r}'). \label{eq:HelmholtzInmogG}
\end{equation}

They are valuated as

\begin{align}
G_{j}(\mathbf{r},\mathbf{r}')&=i\pi H_{0}^{(1)}(k_{0}\sqrt{\epsilon_{rj}}\left|\mathbf{r}-\mathbf{r}'\right|), \label{eq:Green2D} \\
\frac{\partial G_{j}(\mathbf{r},\mathbf{r}')}{\partial\mathbf{n}'_{j}}&=i\pi k_{0}\sqrt{\epsilon_{rj}}\mathbf{n}'_{j}.\frac{(\mathbf{r}-\mathbf{r}')}{\left|\mathbf{r}-\mathbf{r}'\right|}H_{1}^{(1)}(k_{0}\sqrt{\epsilon_{rj}}\left|\mathbf{r}-\mathbf{r}'\right|), \label{eq:DfGreen}
\end{align}
where $H^{(1)}_{0}(\cdotp)$ ($H_{1}^{(1)}(\cdotp)$) denotes the Hankel function of the first class and order zero (one). 

By means of the boundary conditions, Eqns.~(\ref{eq:cc1probSimAxial}-\ref{eq:cc2probSimAxial}), it is possible to decouple the integral equations (\ref{eq:psi0DC}-\ref{eq:psi2DC}), so they will depend only on the fields $\psi_{\alpha}^{(0)}(\mathbf{r'}_{m})$ on each scatterer or contour. Solving numerically for $\psi_{\alpha}^{(0)}(\mathbf{r'}_{m})$ and their normal derivatives over the boundaries $\frac{\partial\psi_{\alpha}^{(0)}(\mathbf{r'}_{m})}{\partial\mathbf{n}'_{m}}$, the fields can be calculated in any region of space. 

\subsection*{Outputs of the method: near and far fields}

\textit{Near fields}. After obtaining the source functions $\psi_{\alpha}^{(0)}(\mathbf{r'}_{m})$ and $\frac{\partial\psi_{\alpha}^{(0)}(\mathbf{r'}_{m})}{\partial\mathbf{n}'_{m}}$, the near fields can be calculated by using the Eqns.~(\ref{eq:psi0DC}-\ref{eq:psi2DC}) together with the boundary conditions (\ref{eq:cc1probSimAxial}-\ref{eq:cc2probSimAxial}). Each pair of functions $\psi_{\alpha}^{(j)}(\mathbf{r'}_{m})$,  $\frac{\partial\psi_{\alpha}^{(j)}(\mathbf{r'}_{m})}{\partial\mathbf{n}'_{m}}$ must be used for each respective contour. \\

\textit{Far fields}. The expression for the far scattered field can be approached from equation \eqref{eq:psi0DC} by making use of the asymptotic approximations of the Hankel functions when $k_{0}\left|\textbf{r}-\textbf{r}'\right|\gg1$.
\begin{eqnarray}
\left.\psi_{\alpha}^{sca}(\textbf{r})\right|_{k_{0}|\mathbf{r}-\mathbf{r}'|\longrightarrow\infty}\equiv\left.\psi_{\alpha}^{sca}(r,\varphi)\right|_{k_{0}|\mathbf{r}-\mathbf{r}'|\longrightarrow\infty} 
\notag\\
=\sqrt{\frac{2}{\pi k_{0}r}}e^{-ik_{0}r+i\frac{3\pi}{4}}\left[T_{1\alpha}(\varphi)+T_{2\alpha}(\varphi)\right],
\end{eqnarray}
In this expression, the distance between cylinders has been supposed to be much smaller than the distance $r$, being $r=|\mathbf{r}|$ the first coordinate of the point $\mathbf{r}=(r,\varphi)$ where the far field is calculated. The functions $T_{m,\alpha}(\varphi)$ ($m=1,2$) are the scattering amplitudes obtained by integral form, defined as 
\begin{eqnarray}
T_{m,\alpha}(\varphi) &= \frac{i}{4}\int_{C_{m}^{'(+)}}ds'\left(ik_{0}(\mathbf{n}'\cdot\textbf{n}_{far}(\varphi))\psi_{\alpha}^{0}(\textbf{r}')+\frac{\partial\psi_{\alpha}^{0}(\textbf{r}')}{\partial\mathbf{n}'}\right) \cdotp \notag \\
& \cdotp e^{-ik_{0}\textbf{n}_{far}(\varphi)\cdot\textbf{r'}}.\label{eq:T-1}
\end{eqnarray}
where the limit $\textbf{n}_{far}(\varphi)=\left.\frac{(\mathbf{r}-\mathbf{r}')}{\left|\mathbf{r}-\mathbf{r}'\right|}\right|_{r\rightarrow\infty}\sim(cos\varphi,sen\varphi)$ has been defined. 

With this formulation, the optical cross sections can be expressed as \cite{VandeHulst}:

\begin{align}
C_{sca,\alpha}&=\frac{2}{\pi k_{0}}\int_{0}^{2\pi}\left|F_{\alpha}(\varphi)\right|^{2}d\varphi,\label{scatt-1} \\
C_{ext,\alpha}&=\frac{4}{k_{0}}Re\left\{F_{\alpha}(\theta=0)\right\} ,\label{ext-1}\\
C_{abs,\alpha}&=C_{ext,\alpha} - C_{sca,\alpha}, \label{Cabs-1}
\end{align}
where $F_{\alpha}(\varphi)=\frac{\sum^{2}_{m=1} T_{m,\alpha}(\varphi)}{|\psi_{\alpha}^{inc}|}$. Notice that the angle of forward scattering, $\theta=\varphi_0+\pi-\varphi$, has been introduced in the argument of $C_{ext,\alpha}$, see Fig.~\ref{fig:1_EsqIntMethod}. In the particular case of having a single wire, all the method can be reduced by setting only one scatterer as $m=1$ above. In particular, $m=1$ only in $F_{\alpha}$ and a radiation pressure's cross section can also be defined through \cite{VandeHulst}
\begin{equation}
C_{pr}=C_{ext}-\left\langle cos\theta\right\rangle C_{sca}. \label{eq:CprGral}
\end{equation}
where the following average must be taken
\begin{equation}
\left\langle cos\theta\right\rangle C_{sca}=\frac{1}{k_{0}^{2}}\int^{2\pi}_0 |F_{\alpha}(\varphi)|^{2}cos\theta d\varphi,
\end{equation}

It is worth mentioning that the method has been implemented and tested to verify the convergence of the integrals. It was considered that the solutions had been converged when the relative error was less than $0.05\%$ between two consecutive discretizations. The method has been subjected to careful testing and extreme situations have been explored, such as $r_{1} \rightarrow 0$, $r_{2} \rightarrow 0$ or $\epsilon_{r1}= \epsilon_{r0}$, $\epsilon_{r2}= \epsilon_{r0}$ or $\mu_{r1}= \mu_{r0}$, $\mu_{r2}= \mu_{r0}$. In both sets of cases, the solution of the problem was naturally reduced to the one corresponding to a dielectric or magnetic solid wire, respectively.

\subsection*{2D Mie Calculations}

In the case of a single wire, the results obtained by the integral method are easily comparable with the results given by the 2D Mie theory \cite{Bohren1998}. Furthermore, the Mie expansion allows identifying the resonant excitations \cite{Abraham2013}. The far fields for each fundamental polarization can be characterized by

\begin{align}
C_{ext,s}&=\frac{4}{k_{0}}Re\left\{ b_{0}+2\sum_{l=1}^{\infty}b_{l}\right\} ,  \label{eq:CextMieS} \\
C_{sca,s}&=\frac{4}{k_{0}}\left(\left|b_{0}\right|^{2}+2\sum_{l=1}^{\infty}\left|b_{l}\right|^{2}\right), \label{eq:CscaMieS} 
\end{align}
and 
\begin{align}
C_{ext,p}&=\frac{2}{x}Re\left\{ a_{0}+2\sum_{l=1}^{\infty}a_{l}\right\} , \label{eq:CextMieP} \\
C_{sca,p}&=\frac{2}{x}\left(\left|a_{0}\right|^{2}+2\sum_{l=1}^{\infty}\left|a_{l}\right|^{2}\right), \label{eq:CscaMieP} 
\end{align}
respectively, where $x=k_0R$ and the following coefficients are defined
\begin{align}
b_{l}&=\frac{\eta_{r10}J_{l}'(\eta_{r10}x)J_{l}(x)-J_{l}(\eta_{r10}x)J_{l}'(x)}{\eta_{r10}J_{l}'(\eta_{r10}x)H_{l}^{(1)}(x)-J_{l}(\eta_{r10}x)H_{l}'^{(1)}(x)}, \label{eq:CoefSCilMieb} \\
a_{l}&=\frac{J_{l}'(\eta_{r10}x)J_{l}(x)-\eta_{r10}J_{l}(\eta_{r10}x)J_{l}'(x)}{J_{l}'(\eta_{r10}x)H_{l}^{(1)}(x)-\eta_{r10}J_{l}(\eta_{r10}x)H_{l}'^{(1)}(x)},  \label{eq:CoefPCilMiea} \\
\end{align}
where $\eta_{r10}=\frac{\sqrt{\varepsilon_{r1}\mu_{r1}}}{\sqrt{\varepsilon_{r0}\mu_{r0}}}$. The electromagnetic resonances for $s$- and $p$-polarizations are related to the complex poles of the coefficients $b_{l}$ and $a_{l}$ respectively. 

Then, in a way analogous to a quantum mechanical problem, a set of integral numbers identify the resonances. For three-dimensional (3D) spheres we deal with three quantum numbers: the radial number $n_r$ and the two angular momentum numbers $l,m$ (do not confuse this $m$ with the number of scatterers given above) \cite{Johnson1993}. For 2D spheres or circular wires, we deal only with two numbers: $n_r$ the radial number and $l$ the azimuthal number. The (magnetic) angular momentum $m$ is ignored since the problem has an ignorable coordinate, e.g. $z$. Differently to the 3D problem of spheres, the angular mode $l=0$ or monopole order can be excited in a dielectric 2D sphere. In spheres, the first allowed mode is $l=1$.

\subsection*{Exact Calculations of Forces and Torques}

Consider the total time-averaged force $\left\langle \mathbf{F}(t)\right\rangle$ exerted by an electromagnetic field to a closed surface $S=S(V)$ that surrounds a regular volume $V$ \cite{Jackson,Novotny}
\begin{equation}
\left\langle \mathbf{F}(t)\right\rangle = \ointop_{S(V)}\left\langle \mathbf{\overleftrightarrow{T}}(\mathbf{r},t)\right\rangle .\mathbf{\hat{n}}da \label{eq:FgralMedio}
\end{equation}
where $\mathbf{\overleftrightarrow{T}}(\mathbf{r},t)$ is the Maxwell stress tensor such that $\left\langle \mathbf{\overleftrightarrow{T}}(\mathbf{r},t)\right\rangle.\mathbf{\hat{n}}=\left\langle\varepsilon_{0}\varepsilon_{r0}(\mathbf{E}.\mathbf{\hat{n}})\mathbf{E}+\mu_{0}\mu_{r0}(\mathbf{H}.\mathbf{\hat{n}})\mathbf{H}-\frac{1}{2}(\varepsilon_{0}\varepsilon_{r0}E^{2}+\mu_{0}\mu_{r0}H^{2})\mathbf{\hat{n}}\right\rangle$ \cite{Stratton}, being $\mathbf{\hat{n}}$ the normal of $S$ pointing to outward direction. $S$ is considered as immersed in a medium of parameters $\varepsilon_{r0},\mu_{r0}$ that cannot support shear stresses \cite{Stratton,Jackson}. This relation for the optical forces is valid only for linear phenomena, and considering the scattering object as rigid \cite{Stratton}.

In a similar way, an expression can be deduced for the net mechanical torque $\mathbf{N}=\frac{d\mathbf{J}_{mech}}{dt}$ acting over the irradiated structure, where $\mathbf{J}=\mathbf{J}_{mech}+\mathbf{J}_{field}$ is the total angular momentum and $\mathbf{J}_{mech}$ -$\mathbf{J}_{field}$- the angular momentum of the matter (field). For a field that satisfies the condition 
\begin{equation}
\left\langle \frac{d}{dt}\mathbf{J}_{field}\right\rangle =0, \label{eq:TorqFieldCond}
\end{equation}
the net time-averaged torque exerted on a rigid arbitrary object inside $S$, is represented by
\begin{align}
\left\langle \mathbf{N}(t)\right\rangle =-\int_{S(V)}\left\langle \overleftrightarrow{\mathbf{T}}
(\mathbf{r},t)\times\mathbf{r}\right\rangle .\mathbf{\mathbf{\hat{n}}}(\mathbf{r})da 
\label{eq:TorqueProm}
\end{align}
If a 2D problem is assumed with axial symmetry with respect to $z$-axis, the expression for the time-averaged force of Eq.~(\ref{eq:FgralMedio}) can be reduced to (SI units)
\begin{eqnarray}
\left\langle d_{z}\mathbf{F}_{\alpha}\right\rangle =\frac{1}{2}Re\Bigg \{ \oint_{C_{l}}\Bigg[\varepsilon_{0}\varepsilon_{r0}\left(\mathbf{E}(\mathbf{r}).\mathbf{\hat{n}}_l\right)\mathbf{E^{*}(\mathbf{r}}) + \nonumber \\
+\left(\mu_{0}\mu_{r0} \mathbf{H}(\mathbf{r}).\mathbf{\hat{n}}_l\right)\mathbf{H}^{*}(\mathbf{r}) + \nonumber \\
-\frac{1}{2}\left(\varepsilon_{0}\varepsilon_{r0}\left|\mathbf{E}(\mathbf{r})\right|^{2}+\mu_{0}\mu_{r0}\left|\mathbf{H}(\mathbf{r})\right|^{2}\right)\mathbf{\hat{n}}_l\Bigg]ds\Bigg \},\label{eq:DensFpromt2D}
\end{eqnarray}
In this case, the radius $\mathbf{r}$ is always contained in the plane $(x,y)$ and the surface integral is reduced to a curvilinear one through $\int_{S_{l}}da=\oint_{C_{l}}ds\int dz$. The contour $C_{l}$ must contain the scatterer $l$ with its own normal $\mathbf{\hat{n}}_l$. In this work, $l=1,2,3$ (see Sec.~\ref{sec:GralResults}); $l=1,2$ correspond to circles closing the wires 1 and 2 respectively and $l=3$ corresponds to a circle closing the entire dimer. As infinite cylinders are assumed, $\int dz\rightarrow\infty$ and a net density of force $d_{z}\textbf{F}_{\alpha}$ must be defined instead of the force $\textbf{F}_{\alpha}$ itself. Similarly, a net density of torque $d_{z}\textbf{N}_{\alpha}$ must be defined instead of torque, and the following equation is deduced from \eqref{eq:TorqueProm} 

\begin{eqnarray}
\left\langle d_{z}\mathbf{N}_{\alpha}\right\rangle = \frac{1}{2}Re\Bigg \{ \oint_{C_{l}}\Bigg[\left(\varepsilon_{0}\varepsilon_{r0}\mathbf{E}(\mathbf{r}).\mathbf{\hat{n}}_l\right)\mathbf{r}\times\mathbf{E^{*}(\mathbf{r}}) + \nonumber \\
+ \left(\mu_{0}\mu_{r0}\mathbf{H}(\mathbf{r}).\mathbf{\hat{n}}_l\right)\mathbf{r}\times\mathbf{H^{*}(\mathbf{r}}) + \nonumber \\
-\frac{1}{2}\left(\varepsilon_{0}\varepsilon_{r0}\left|\mathbf{E}(\mathbf{r})\right|^{2}+\mu_{0}\mu_{r0}\left|\mathbf{H}(\mathbf{r})\right|^{2}\right)\mathbf{r}\times\mathbf{\hat{n}}_l\Bigg]ds\Bigg \}. \label{eq:TorqueProm2D}
\end{eqnarray}
where $d_{z}\mathbf{N}_{\alpha}=d_{z}N_{\alpha}.\mathbf{\hat{z}}$ has dimensions of force. In the case of \textit{spin} torques, the last term of the Eq.~(\ref{eq:TorqueProm2D}) can be eliminated if the contour $C_{l}$ is simplified to a circle such that $\mathbf{r}\times\mathbf{\hat{n}}_l=0$.

Notice that for plane waves, the condition \eqref{eq:TorqFieldCond} is automatically satisfied. Notice also that the Eqns.~(\ref{eq:DensFpromt2D}-\ref{eq:TorqueProm2D}) are valid for both fundamental polarizations. In particular, the first (second) term of these equations always cancels out under for polarization $s$ ($p$) because of the orthogonality of the fields.

In the case of a single wire, the force density or radiation pressure density can be calculated also as \cite{Bohren1998} 
\begin{align} 
\left\langle d_{z}\mathbf{F}_{\alpha}\right\rangle = \frac{I_0}{c}C_{pr}\frac{\mathbf{k}_0}{k_0} \label{eq:RadPressDzF}
\end{align}
 where $C_{pr}$ is given by the Eqn.~(\ref{eq:CprGral}) and $I_0$ is the intensity of the incident wave \cite{VandeHulst}. In the case of using the 2D Mie's formulation, $C_{pr}$ can be calculated by taking the Eq.~(\ref{eq:CprGral}) with the optical cross sections given by the Eqns.~(\ref{eq:CextMieS})-(\ref{eq:CscaMieS}) and (\ref{eq:CextMieP})-(\ref{eq:CscaMieP}) for each polarization, respectively. To ensure the coherency of the methodology used here, all the spectra for single wires have been calculated by using both integral and Mie formulations. Both methods have given the same results (not shown here). 
 
 An electric field amplitude $\left|\mathbf{E}^{inc}\right|$ was assumed to calculate normalized mechanical densities. These densities will be by the factor $4\pi\varepsilon_{0}\left|\mathbf{E}^{inc}\right|^2$. In this way, the results are valid to use arbitrary intensities of illumination. However, when using big incident powers or relatively long times in making measures, heating effects should be included in the problem \cite{HeatTransfQuidant,HeatTransf2D}. In practice, pulsed lasers can be used to limit heating in the particles. In this paper, neither the thermal fluctuations due to radiative heat transfer nor pulsed lasers are considered in the response.

In the following section, the embedding medium will be assumed to be the air or vacuum, i.e. $\varepsilon_{r0}=1$, $\mu_{r0}=1$, and the systems are assumed to be made by silicon so that $\varepsilon_{r1}=\varepsilon_{Si}$ and $\mu_{r1}=1$ or $\varepsilon_{r1}=\varepsilon_{r2}=\varepsilon_{Si}$ and $\mu_{r1}=\mu_{r2}=1$ for isolated wires or dimer systems, respectively.

\section{Results \label{sec:GralResults}} 

As it was explained above, the forces and torques will be shown as scaled magnitudes. The force densities will give the same units as the optical cross sections in two dimensions, i.e. in units of $nm$. Then, they are fully comparable to the observables of the system in the far-field region. The 2D torques, as given in $nm^2$, account as an effective 2D volume of action of the induced forces.

The system of an isolated silicon wire is studied first in order to explore the influence of the MDRs on the forces. The results are analyzed in terms of 2D Mie expansions to identify the MDRs as multipolar contributions. This part is fundamental to understand the optical response induced in the dimer system given further below.

\subsection*{Preliminary Results: Isolated wires}

\subsubsection*{Far-field properties}

\begin{figure*}[!h]
 \begin{centering}
 \includegraphics[width=15cm,height=15cm,keepaspectratio]{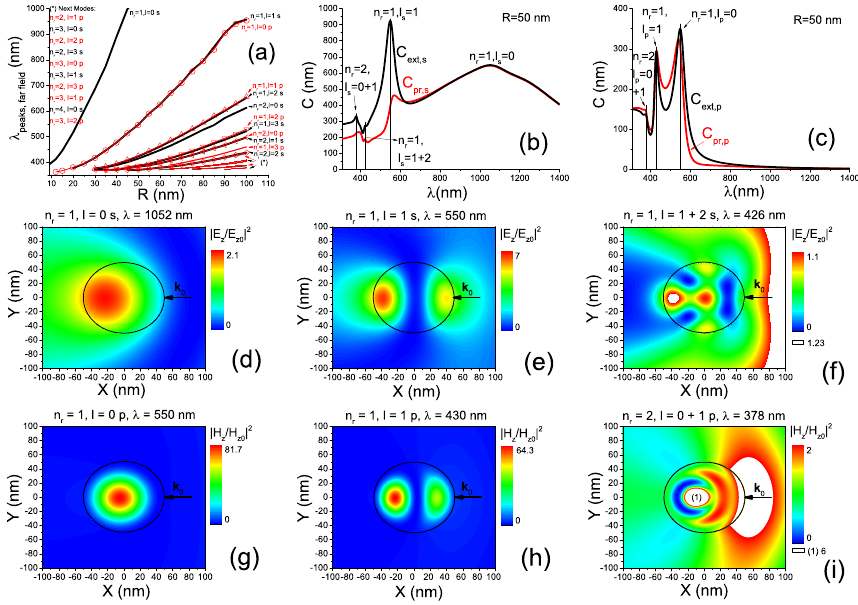} 
 \par\end{centering}
 \caption{\label{fig:2_Si}  Optical properties of Si isolated nanowires and influence of the MDRs on the optical forces. (a) Dispersion curves of the modes as a function of the radius $R$ of the wire (black -red- for $s$ -$p$- polarization in the online version). Panels (b)-(i) for the case $R=50$ nm of (a); (b) -(c)- Comparison of spectra of Radiation's pressure cross section vs. extinction under $s$ -$p$- polarized waves. The vertical lines serve to compare the spectral locations of the MDRs. (d)-(i) Maps of the near and inner fields at the location of the first resonances; the small excitation at $\lambda=378$ nm is not shown. (d)-(f) under $s$-polarization. (g)-(i) under $p$-polarization. The saturated color scale is labeled by the reference number which shows the field intensity in the regions in white. As a reference, the arrow on each panel indicates the direction of the incident wave.}
\end{figure*}

The Fig.~\ref{fig:2_Si} illustrates the appearance of the MDRs in realistic silicon wires and its influence on the induced optical forces. An evolution of modes can be seen for a silicon nanowire as a function of the radius $R$, see \ref{fig:2_Si}(a). The detailed calculations shown in \ref{fig:2_Si}(a) offer a general overview of the MDRs expected for the spectra of radiation pressures at one particular value of radius $R$. Wires with radii equal or near the value $R=50$ nm are chosen in this work as examples for exhaustive studies of forces. Panels \ref{fig:2_Si}(b) and (c) show the spectra for the case $R=50$ nm of \ref{fig:2_Si}(a), for each fundamental polarization respectively. Curves in black (red) line in \ref{fig:2_Si}(b) and (c) correspond to the extinction's (radiation's pressure) cross section for each polarization. All the results of the Fig~\ref{fig:2_Si} have been calculated by the integral method. However, the MDRs have been identified with labels obtained by 2D Mie calculations up to the $l=3$ angular orders. The curves in \ref{fig:2_Si}(a) have been shown before in \cite{Abraham2013} and the integral method had already been compared against Mie theory, giving very accurate results. Notice the appearance of several modes for relatively small wires in the subwavelength scale. This is possible due to the high values of the permittivity given by the dielectric function of silicon. The dispersion curves of the MDRs do not result in straight lines because the permittivity is a spectral function. An empirical rule $l_p=l_s-1$ can also be appreciated for the first modes that appear in the spectra. With the exception of the first MDR for $s$-polarization, the locations of the MDRs in the spectra are very similar for both polarizations.

A fast growing can be observed for the curve $l=0$ of the $s$-mode when $R$ is increased, see \ref{fig:2_Si}(a). This fact means that the mode $l_s=0$ can be tuned up to very long wavelengths even for wires with relatively small radii. In general, this MDR is seen to have a very different behavior from the rest of the modes in near- and far-field regimes. A more detailed study about this special mode can be found, for instance, in Ref.~\cite{VanBladel1977}.

\subsubsection*{Radiation Pressure}

The structure of the spectra of radiations' pressures and the extinctions are similar in \ref{fig:2_Si}(b) and (c) \cite{Abraham2013}, with the exception of relatively small wavelength shifts in the peaks of the excitations \cite{Menzel2014,Gutierrez2016,Yuffa2016}. Besides, the silicon dielectric function has a strong imaginary part at short wavelengths \cite{Abraham2013} that influences the shifts and the relative intensities of the force peaks with respect to the far-field observable in the high-energy region \cite{Menzel2014}. The absorption process is taken into account in the balance of the linear momentum transferred to the wire. The radiation pressure cross section gives the optimal length (2D area) to exert force by light on the wire. This mechanical magnitude is affected by the MDRs induced on the structure. 

For the case of $R=50$ nm, four MDRs can be distinguished in the $s$-spectrum of \ref{fig:2_Si}(b) and three MDRs in the $p$-spectrum of \ref{fig:2_Si}(c). Although the curves have overlapped peaks, it can be said that roughly the same peak resolution as the far-field magnitudes is obtained by the spectra of forces for single wire (scattering and absorption cross sections are not shown here). The resolution of the spectral magnitudes will be compared with those for coupled systems further below. In particular, notice the structure of the first MDR corresponding to the mode $l=0$ in the pressure spectra under $s$-polarization: the excitation is very wide and strongly overlaps with the rest of the MDRs. This behavior does not occur under $p$-polarization and it is a common characteristic of all the spectral curves shown in this work.

As the integral method takes the force results from the near fields, one can say that these results may be expected to be different than those obtained by the Eqns.~(\ref{eq:RadPressDzF}) and (\ref{eq:CprGral}), a method which involves the far-fields. However, no difference can be appreciated between both methods. The near fields carry information about the contributions due to the evanescent waves but these have propagation constants parallel to interfaces of the scatterer \cite{Abraham2015_1}. The circulation of such waves around the wire's surface cannot account for new contributions to the radiation pressure as the field inductions result highly symmetric. The anticlockwise circulation of such waves is equal to the clockwise circulation because the incident (propagating) field is homogeneous. Then, there is no difference in using both the near-field or the far-field method when dealing with the force exerted by a plane wave on a single wire.

Of course, there is no optical torque at all for single wire because the highly symmetric scatterer is being ``pushed'' by plane waves with linear polarization. The wire region responds with fields that result symmetric with respect to the direction of the illumination under both polarizations, see the field maps \ref{fig:2_Si}(d-i) around the wire. 

\subsubsection*{Near and inner fields}

All the far-field features can be related with the field distributions of the MDRs around the wire, i.e. near and inner fields. Here, the maps \ref{fig:2_Si}(d-i) illustrate the relevant field structures around the wire of $R=50$ nm of \ref{fig:2_Si}(b) and (c). The panels \ref{fig:2_Si}(d-f) show the behaviour of the wavefunctions for the first three MDRs than can be appreciated from the far-fields under $s$-polarization. Similarly, the panels \ref{fig:2_Si}(g-i) show the behavior at the location of the three MDRs that can be distinguished under $p$-polarization. The spectral locations of the MDR are taken from the far-field curves; the shifts that can occur between the resonances at near and far fields were shown to be small, see \ref{fig:2_Si}(b-c). The radial and angular distribution of the fields are characterized by the integral numbers given by the Mie expansion; $n_r$ corresponds to the number of maxima of the field intensity along the radial direction. On the other hand, $2l$ corresponds to the number of maxima along the azimuthal angle from $0$ to $2\pi$ rad. 

Observe the structure of the fields under $s$-polarization, panels \ref{fig:2_Si}(d-f). Notice that the wavefunction, the electric field in this case, is not confined to the wire region. The field distribution is spread to another bigger region as if it were confined to the inside of a scatterer which size is bigger than the real one. On the contrary, under $p$-polarization, panels \ref{fig:2_Si}(g-i), the wavefunction is the magnetic field and it is strongly confined into the wire region or the region where the high-contrast medium exists. Thus in general, we will see that the $p$- or magnetic modes are much more confined and more enhanced than the $s$- or electric modes.

The near-field structures of the MDRs are seen to be well characterized by their radial and polar modulations. Although the MDRs have an intrinsic bulk nature, they can eventually induce some surface concentration of the fields, as seen in \ref{fig:2_Si}(i). This effect is a result of the interference of the scattered and the incident fields. Another result of this interference is the clear ``blowing'' effect that can be seen in the patterns. The incident fields distort the near-field structures as if they were blown to the forward direction with respect to the illumination.

It is worthwhile to notice the huge enhancement of field intensity obtained for the first MDRs under $p$-polarization \ref{fig:2_Si}(g-h). This behavior will play a key role in the forces and torques induced on the systems of coupled wires.

The maybe trivial behavior of the MDRs shown on the forces induced on single wires will not prevail for systems of coupled wires. The complex interaction between the wires will be manifested on the coupling of MDRs in dimer systems. The scheme of hybridization of MDRs is not simple and subsequently entails unusual properties on the optomechanical inductions.

\subsection*{Homodimers \label{sub:HD}}

The homodimer system will be explored only under illumination with $\varphi_0=90$ deg. A more comprehensive study of the inductions for the two relevant angles $\varphi_0=0;90$ deg will be carried out in the following section devoted to heterodimers. The coupled wires present hybridized resonances with respect to the resonances of the isolated wires. The induced forces and torques are presented by means of a particular example for a gap of $d=5$ nm between the wires. 

\subsubsection*{Far-field properties}

\begin{figure*}[!h]
 \begin{centering}
 \includegraphics[width=15cm,height=15cm,keepaspectratio]{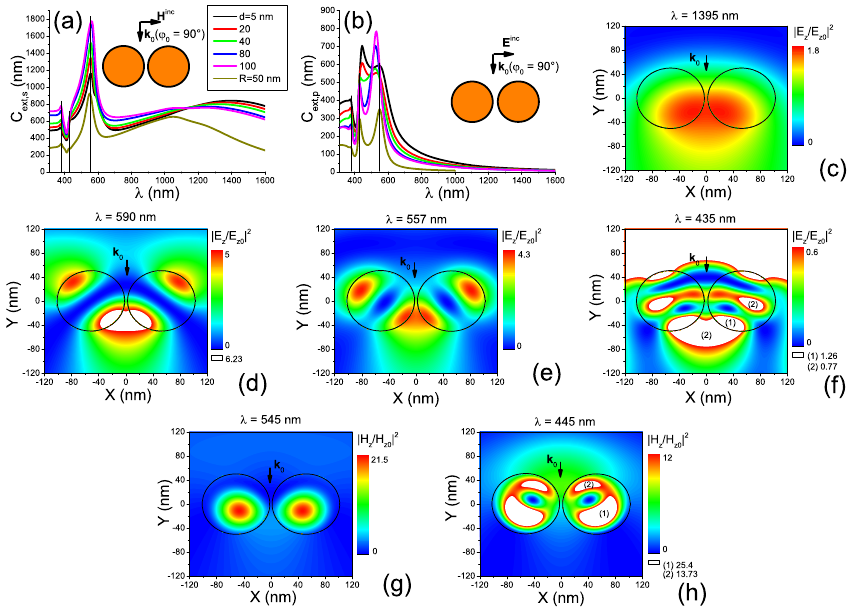} 
 \par\end{centering}
 \caption{\label{fig:3_HomoD}  Far-field and near-field responses by Si homodimers of nanowires of radii $r_1=r_2=50$ nm. The angle of illumination is $\varphi_0=90$ deg. (a) and (b) Spectra of extinction curves as a function of the gap $d$ under polarization $s$ and $p$ respectively. The vertical lines serve to compare the spectral locations of the MDRs. The curves in dark yellow line correspond to the response by the isolated wire of $R=50$ nm or $d\longrightarrow\infty$. (c)-(f) Near-field maps at the position of resonances for the cases $d=5$ nm of (a-b). (c)-(g) under $s$-polarization. (g)-(h) Under $p$-polarization. The saturated color scale is labeled by reference numbers which show the field intensity in the regions in white.}
\end{figure*}

The far- and near-field responses for the homodimer are shown in Fig.~\ref{fig:3_HomoD}, panels (a-b) and (c-h) respectively. Notice the behavior of the extinction cross sections when the gap is varied; the response of the single wire $R=50$ nm is also shown for comparison (curve in dark yellow line in online version; see \ref{fig:3_HomoD}(a) and (b)). Although very different gaps have been calculated, the MDRs remain almost at the same spectral locations with the exception of the first mode (the one with the lowest energy) of the dimer under $s$-polarization. This mode presents a great shift of the spectral location with the gap if compared against $n_r=1,l_s=0$ of the single wire; see the previous results for details. Although there is no realistic hybridization model for dielectric wires, it is easy to speculate that the dimer modes can be built with multiple multipolar resonances, in a manner similar to that found for metallic dimers \cite{NordlanderPHDim}. This hybridization has been realized for silicon dimers of 3D spheres \cite{Albella2013,Zywietz2015,Bakker2015}. The photonic transitions $\Delta l\neq0$ are characteristic of the dimer geometry by the series expansions of the eigenfunctions of each wire over the functions of the other wire \cite{Jackson}. In this way, one can argue that the modes $n_{rm}=1,l_{sm}=0$ of single wire for $m=1,2$ must be important in the hybridization of the first $s$-mode as they conserve a similar spectral structure, see \ref{fig:3_HomoD}(a). However, as this MDR results very spread in all the spectra, their excitations are not so relevant in the system's characterization. The rest of the MDRs remain quite static in the spectra as the gap is varied for the two polarizations. Actually, the spectra for the homodimer is very similar to the spectra for the single wire under both polarizations. This is due to the nature of the MDRs as volume resonances. The fields at these resonances are very confined to the wires' region and the spectral curves are not so sensitive to the geometric variations of the dimer, with the exception of the shifts in the spectral locations of the first MDR. On the other hand, there are strong variations of the intensities of the MDRs with the gap, as expected. In particular, for the shortest gap, the extinction for $s$-polarization reaches maxima of almost twice the reached maxima of the extinction for $p$-polarization, see the second peaks in \ref{fig:3_HomoD}(a) and (b).

\subsubsection*{Near and inner fields}

The maps of near and inner fields of \ref{fig:3_HomoD}(c-h) help us to give a better idea of the hybridizations of MDRs that occur in the homodimer. For the sake of clarity, the most relevant maps are shown here, the highest-energy modes at $\lambda\simeq 375-380$ nm are not shown. Then, four modes are shown in \ref{fig:3_HomoD}(c-f) for $s$-polarization. Similarly, two MDRs are shown in \ref{fig:3_HomoD}(g-h) for $p$-polarization. In general, a great coupling of the fields that resonate at each wire can be seen in the maps. However, as said, the inner structure of the MDRs can be observed to be similar to those found for the single wire. The map \ref{fig:3_HomoD}(c) appears to support the hypothesis of the hybridization of the modes $n_{rm}=1,l_{sm}=0$. Similarly, \ref{fig:3_HomoD}(g) shows an analog behaviour for $p$-polarization, i.e. a hybridization of the modes $n_{rm}=1,l_{pm}=0$. The maps \ref{fig:3_HomoD}(d-e) show a strong influence of the modes $n_{rm}=1,l_{sm}=1$ of the single wire, which open up to both sides due to the effect of the illumination and the multiple scattering. The panel \ref{fig:3_HomoD}(h) shows a similar behavior for $p$-polarization but, in this case, the structures close up to the inner region between the wires. Again, a big field concentration is obtained under $p$-polarization. Notice the values of the intensity reached in \ref{fig:3_HomoD}(g-h) which are several times the value given by the incident wave.

\subsubsection*{Homodimers' forces}

The methodology for calculating the mechanical magnitudes is the same than the presented in Refs~\cite{Abraham2016,Abraham2018Ag} in agreement with the general methodology introduced in the previous section, \ref{sec:Methods}. The resultant curves use the same color code than the circles of integrations; black for wire $1$ or left wire closed by $C_1$, red for wire $2$ or right wire closed by $C_2$ and green for the whole system closed by $C_3$. It worthwhile to clarify that the curve $C_3$ can be arbitrary while it closes properly the entire system; it could be a circle as well as any other curve and it would give the same results for the calculations of mechanical magnitudes. In the following figures, the curves $C_3$ are drawn as ellipses in the inset schemes for convenience but, in fact, they were used as circles in the calculations.

\begin{figure*}[!h]
 \begin{centering}
 \includegraphics[width=15cm,height=15cm,keepaspectratio]{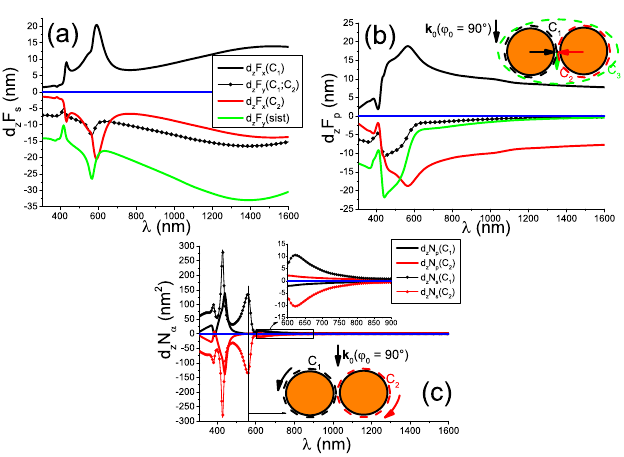}
 \par\end{centering}
 \caption{\label{fig:4_HomoD_Fcs}  Optical forces and torques exerted on Si homodimers of radii $r_1=r_2=50$ nm and $d=5$ nm under plane-wave illumination with linear polarization. The incident angle is $\varphi_0=90$ deg. (a) and (b) Spectra of density of forces under polarization $s$ and $p$ respectively. (c) induced torques under both polarizations. The inset graphic shows a zoom corresponding to the window $(-15,15)$ nm$^2$ in the zone $\lambda=(600,900)$ nm. The zeros of the scales are highlighted by blue lines.}
\end{figure*}

After a careful relation between the far-field spectra and the induced optical forces, i.e. \ref{fig:3_HomoD}(a-b) vs. \ref{fig:4_HomoD_Fcs}(a-b), the MDRs can be identified and they appear in the mechanical observables -see also \ref{fig:4_HomoD_Fcs}(c)-. Even the highest-energy mode can be seen to appear lightly in the forces around $\lambda\simeq375-382$ nm. Under the configuration $\varphi_0=90$ deg, the force components of scattering and binding can be distinguished, see in \ref{fig:4_HomoD_Fcs}(a-b): curves in black dotted line and in green solid line are scattering components while they curves in black and red solid lines are binding components.  Both sets of curves carry the information of the interaction between the wires. Of course, the system is pushed ``down'' as a whole in $-y$ direction by radiation pressure due to the incident waves (green curves in \ref{fig:4_HomoD_Fcs}(a) and (b)). The black dotted curve represents two equal contributions for the induced density of force along the $y$-axis, each one corresponding to each wire respectively. This curve logically is half of the green curve for the induced force for the system, i.e. $d_zF_y(C_3)=d_zF_y(C_1)+d_zF_y(C_2)=2d_zF_y(C_1)=2d_zF_y(C_2)$. To estimate the effect of the coupling on the radiation pressure, one can compare the curves of the scaled $y$-forces of \ref{fig:4_HomoD_Fcs}(a-b) with the scaled force for single wire as $\frac{|\left\langle d_{z}\mathbf{F}_{\alpha}\right\rangle|}{4\pi\varepsilon_{0}\left|\mathbf{E}^{inc}\right|^2} = \frac{1}{8\pi}C_{pr}$ obtained by the Eq.~(\ref{eq:RadPressDzF}), where $C_{pr}$ is given in panels \ref{fig:2_Si}(b-c). The $y$-forces have the same order of magnitude than the radiation pressure for the single wire.

It is worth to notice that the binding forces show an attractive behavior under both polarizations in all the spectra -see the inset illustration on panel \ref{fig:4_HomoD_Fcs}(b)-. This means that all the hybridized MDRs studied in Fig.~\ref{fig:3_HomoD} for homodimers result in bonding modes for this particular geometric configuration. This may be simple dynamics will not be held for heterodimers' configurations, as it will be seen further below.

\subsubsection*{Homodimers' torques}

As it was studied for metallic dimers \cite{Abraham2016,Abraham2018Ag}, unexpected torques are induced by linear polarization as a result of the realistic interaction between the silicon wires, panel \ref{fig:4_HomoD_Fcs}(c). In this case, due to the nature of the MDRs, spin torques can be seen under the illumination with both polarizations $s$ and $p$; see the curves in dotted line vs. the curves in solid line respectively in \ref{fig:4_HomoD_Fcs}(c). Although no net induced torque exists for the system, as expected by the homodimer's symmetry, net spin torques exist for each wire and they appear in \textit{coordinated} form \cite{Abraham2016}. Observe that the red and black curves in \ref{fig:4_HomoD_Fcs}(c) are always equal but opposite in such a form they are always balanced to zero for the whole system. As it was expressed previously \cite{Abraham2016,Abraham2018Ag}, the spectral spins have in general more resolution than the optical forces. The peaks corresponding to the MDRs are narrower and less overlapped than those found in the spectra of forces. This conclusion is now of general validity in a way independently of the materials of the wires and it will be also concluded for the heterodimers' case. The torques are more suitable observables of the system, and they are preferable also when choosing a near-field observable of the wires' interaction. Even more, the torques have information of the interaction that is not included in the induced forces. The signs of the spin torques at the spectral locations of the MDRs is one example of this. While the induced binding forces have the same sign in all the spectra of both polarizations $s$ and $p$, the torques have not. In this way, the definition of bonding and antibonding of the modes need to be revised as it was pointed out in \cite{Abraham2018Ag} for metallic dimers.

In the spirit of the previous works, a new relation between observables of both near and far-fields can be realized. With this hypothesis at hand, the maps of the Fig.~\ref{fig:3_HomoD} can be examined in spite of an explanation for the torques found in silicon dimers. Remarkably, there is a connexion between the torques and the presented maps because there are no induced torques for the first MDRs under both $s$- and $p$-polarizations, see \ref{fig:4_HomoD_Fcs}(c). The first MDR is expected for $s$-polarization for $\lambda\approx 1395-1400$ nm, see \ref{fig:3_HomoD}(c) and there is no induced torque for long wavelengths in \ref{fig:4_HomoD_Fcs}(c). Similarly, the same happens under $p$-polarization; the first MDR is expected to appear around $\lambda\approx545$ nm but this resonance does not appear in the spectra of the induced torques, see panel \ref{fig:3_HomoD}(g) and panel \ref{fig:4_HomoD_Fcs}(c) with its inset graphic. The reason for this feature is that the two near-field patterns at these locations are highly symmetric while the other maps clearly show symmetric but bent orientations. These stationary orientations have specular symmetric with respect to the $y$-axis or the illumination direction but show a preferable angle, see panels \ref{fig:3_HomoD}(d-f) and (h). These asymmetries support a reason for the spin torques to exist in the silicon homodimers. Some kind of symmetry breaking is induced by the photonic interaction itself and it would produce the spins. Meanwhile, the gradient of the field distributions in the gap between the wires seems to play a role in bringing the wires together. 

Although a geometric, natural symmetry breaking exists for heterodimers, a similar asymmetric field induction will happen that originates also spin torques. In addition, the dissimilar wires will originate orbital torques. These effects are seen further below in the following section.

\subsection*{Heterodimers \label{sub:HeteroD}}

In this section, an example of parameters $r_1=50$ nm, $r_2=40$ nm and $d=5$ nm is taken as an illustration of the unusual properties of silicon heterodimers. The MDRs can change due to both geometric changes or changes in the relative incident angles as well as due to polarization states. Here, a study is realized for the two relevant directions $\varphi_0=0;90$ deg. The reason for including the configuration $\varphi_0=0$ deg will be clear when the mechanical results are analyzed. As above, the results in far and near fields are related to the mechanical results.

\subsubsection*{Far-field properties}

\begin{figure*}[!h]
 \begin{centering}
 \includegraphics[width=15cm,height=15cm,keepaspectratio]{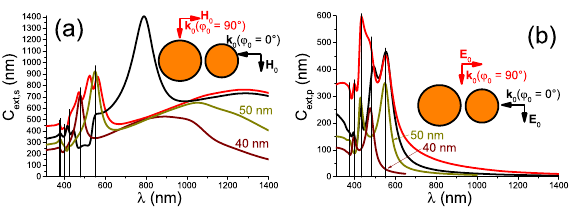}
 \par\end{centering}
 \caption{\label{fig:5_HeteroD_FF}  Far-field response by Si heterodimer of radii $r_1=50$ nm, $r_2=40$ nm and $d=5$ nm. (a) and (b) under polarization $s$ and $p$ respectively. The angles examined are $\varphi_0=0;90$ deg, curves in black and red line in the online version. The curves in dark yellow and brown lines correspond to the response by the isolated wires of $R=50$ nm and $R=40$ nm, respectively. The vertical lines serve to compare the spectral locations of the MDRs.}
\end{figure*}

The far-field responses at $\varphi_0=0;90$ deg are compared between themselves for both polarizations $s$ and $p$ in Fig.~\ref{fig:5_HeteroD_FF}. The spectra for the cases $R=50$ nm and $R=40$ nm of isolated wires have been also added for comparison. The differences of the heterodimer's spectra with the cases $R=50$ nm and $R=40$ nm of isolated wires already indicate a possible hybridization scheme for the MDRs of the heterodimer. The coupling between the wires changes the energies of the MDRs of the system. The vertical lines drawn in the panels help to establish possible relations between the MDRs of the different spectra. 

The first MDR for the heterodimer under $s$-polarization is shown to be very sensitive to the coupling if compared with the first MDRs of the curves for isolated wires. On the contrary, poor sensitivity is found for the MDRs of the system under $p$-polarization; the MDRs seem to have almost the same spectral locations than the MDRs for the isolated wires. For the two fundamental polarizations, both spectra $\varphi_0=0;90$ deg show almost the same spectral location for the first MDR. 

On the other hand, the extinction under illumination $s$ with $\varphi_0=0$ deg is recognized for a strong excitation at $\lambda=790$ nm, see \ref{fig:5_HeteroD_FF}(a). In practice, this MDR can be excited by means of an illumination with a Ti: sapphire laser tuned to this wavelength \cite{Sule2017}. There is no other MDR in this spectral region for illumination with $\varphi_0=90$ deg.

The peaks that appear in the heterodimer's spectra at a similar location to the peaks of the curve for the wire of $R=40$ nm may be characteristic of the dissimilar wires. That is, those MDRs allow us to identify the heterodimer nature of the system. In particular, those MDRs can be discriminated for the spectra $s$ and $p$ for  angle $\varphi_0=90$ deg; and those peaks lie in regions of high energies, e.g. $\lambda\simeq475$ nm for the red curve in \ref{fig:5_HeteroD_FF}(a) and $\lambda\simeq400;470$ nm in \ref{fig:5_HeteroD_FF}(b). The rest of the MDRs in the spectra $s$ and $p$ lie in the close spectral locations between them, with the exception of the peaks at $\lambda\simeq525;565$ nm in the curve in red line that seem to be a splitting of the peak $\lambda=550$ nm of the curve in dark yellow line for the isolated wire $R=50$ nm. This may be a result of the hybridization scheme for this particular example of silicon heterodimers. On the other hand, all the curves shown have very different intensities of excitation of the MDRs. This phenomenon is another consequence of the coupling of the wires and could be properly described by an adequate hybridization model.

\subsubsection*{Near-field properties}

\begin{figure*}[!h]
 \begin{centering}
 \includegraphics[width=15cm,height=15cm,keepaspectratio]{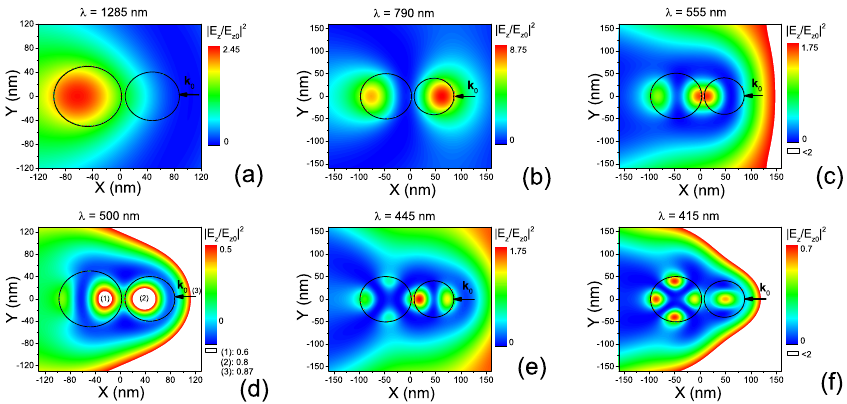} 
 \par\end{centering}
 \caption{\label{fig:6_HeteroD_NFs0}  Near-field maps of the modes of the Si heterodimer of radii $r_1=50$ nm, $r_2=40$ nm and $d=5$ nm under $s$-polarization. The incident angle is $\varphi_0=0$ deg. The saturated color scale is labeled by reference numbers which show the field intensity in some regions in white.}
\end{figure*}

\begin{figure*}[!h]
 \begin{centering}
 \includegraphics[width=15cm,height=15cm,keepaspectratio]{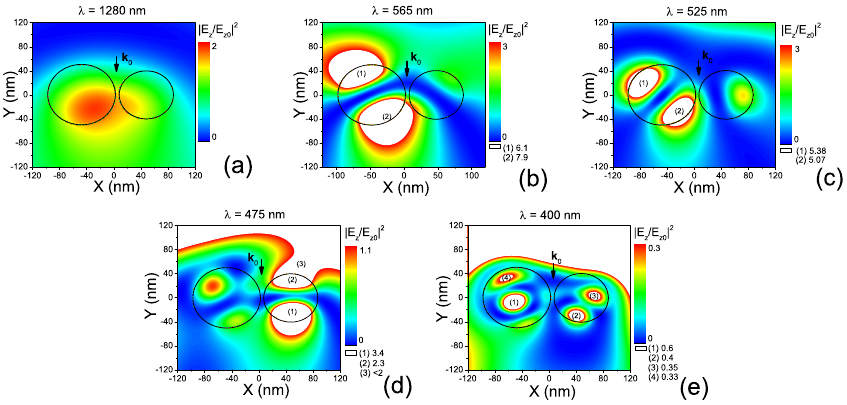}
 \par\end{centering}
 \caption{\label{fig:7_HeteroD_NFs90}  Near-field maps of the modes of the Si heterodimer of radii $r_1=50$ nm, $r_2=40$ nm and $d=5$ nm under $s$-polarization. The incident angle is $\varphi_0=90$ deg. The saturated color scale is labeled by reference numbers which show the field intensity in some regions in white.}
\end{figure*}

\begin{figure*}[!h]
 \begin{centering}
 \includegraphics[width=15cm,height=15cm,keepaspectratio]{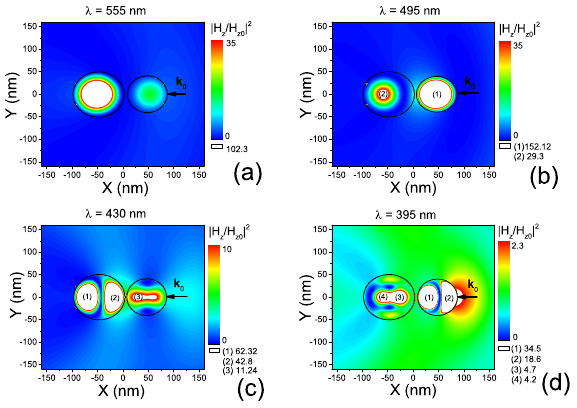}
 \par\end{centering}
 \caption{\label{fig:8_HeteroD_NFp0}  Near-field maps of the modes of the Si heterodimer of radii $r_1=50$ nm, $r_2=40$ nm and $d=5$ nm under $p$-polarization. The incident angle is $\varphi_0=0$ deg. The saturated color scale is labeled by reference numbers which show the field intensity in some regions in white.}
\end{figure*}

\begin{figure*}[!h]
 \begin{centering}
 \includegraphics[width=15cm,height=15cm,keepaspectratio]{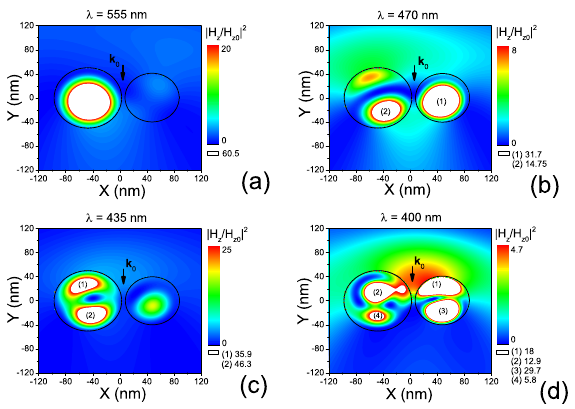}
 \par\end{centering}
 \caption{\label{fig:9_HeteroD_NFp90}  Near-field maps of the modes of the Si heterodimer of radii $r_1=50$ nm, $r_2=40$ nm and $d=5$ nm under $p$-polarization. The incident angle is $\varphi_0=90$ deg. The saturated color scale is labeled by reference numbers which show the field intensity in some regions in white.}
\end{figure*}

To gain knowledge of the far-field behavior of the coupling between the wires, the relevant near-field maps are shown. They illustrate the spectral locations of the first MDRs (lowest-energy modes), see Figs.~\ref{fig:6_HeteroD_NFs0}-\ref{fig:9_HeteroD_NFp90}. The maps of Figs.~\ref{fig:6_HeteroD_NFs0}-\ref{fig:7_HeteroD_NFs90} correspond to the incident angles $\varphi_0=0;90$ deg respectively under $s$-polarization while the maps of Figs.~\ref{fig:8_HeteroD_NFp0}-\ref{fig:9_HeteroD_NFp90} correspond to $\varphi_0=0;90$ deg respectively under $p$-polarization. In general, it is easy to see how several hybridized MDRs are entering in the spectra when the energy is growing, Fig.~\ref{fig:5_HeteroD_FF}. The maps for heterodimers show clearly the symmetry breaking in the structure by means of asymmetric field patterns. Logically, the patterns under the configuration $\varphi_0=0$ deg hold the symmetry with respect to the $x$-axis while the patterns under illumination with $\varphi_0=90$ deg do not hold any symmetry, i.e. symmetry is broken with respect to both $x$- and $y$-axes. 

The monopolar modes for the whole structure appear in \ref{fig:6_HeteroD_NFs0}(a) and in \ref{fig:7_HeteroD_NFs90}(a) under $s$-polarization in a way analogous to the mode found in \ref{fig:3_HomoD}(c) for $\varphi_0=90$ deg. Let us allow to call it monopolar mode in analogy with the mode $n_r=1$,$l_s=0$ of a single wire, this definition is also used in the nomenclature of the molecular theory \cite{Demtroder}. In particular, the maximum field distribution occurs around the wire $1$ for \ref{fig:6_HeteroD_NFs0}(a) as if the illumination would blow the scattered field to the left. In addition, the effective wavelength $\lambda_{eff}=\lambda/\sqrt{\epsilon_{Si}}$ of this energy is too long to include a modal structure inside the wire $2$. On the other hand, the patterns of the maps of \ref{fig:3_HomoD}(c) and \ref{fig:7_HeteroD_NFs90}(a) are very similar, but this latter one results asymmetric with respect to the $y$-axis due to the geometric symmetry breaking of the dimer.

The strongest MDR that appears in \ref{fig:5_HeteroD_FF}(a) at $\lambda=790$ nm under illumination $s$, $\varphi_0=0$ deg, corresponds to the field pattern of \ref{fig:6_HeteroD_NFs0}(b). Interestingly, the scaled field intensity reaches around nine times the intensity of the incident field. This map structure of the fields seems not allowed to exist under $s$-polarization and $\varphi_0=90$ deg, see maps of Fig.~\ref{fig:3_HomoD} for homodimers' resonances and Fig.~\ref{fig:7_HeteroD_NFs90} for heterodimer's resonances. It is a characteristic mode of the configuration $\varphi_0=0$ deg. The pattern resembles a mode like $n_r=1$, $l_s=1$ for isolated wire as if it were the response by a bigger wire corresponding to the entire system. However, the structure could also be seen as a hybridization of two monopolar modes of isolated wires. The rules of hybridization are not affordable with realistic wires.

The next two patterns (c) and (d) of the Fig.~\ref{fig:6_HeteroD_NFs0} resembles hybridizations using the first MDRs of single wires. Pattern \ref{fig:6_HeteroD_NFs0}(c) appears to be built with combinations like $n_{rm}=1$, $l_{sm}=1$ and \ref{fig:6_HeteroD_NFs0}(d) like $n_{r1}=1$, $l_{s1}=1$ with $n_{r2}=1$, $l_{s2}=0$.  Panels \ref{fig:6_HeteroD_NFs0}(e) and (f) could show molecular-like hybridizations of $n_{r1}=1$, $l_{s1}=2$ with $n_{r2}=1$, $l_{s2}=1$ and with $n_{r2}=1$, $l_{s2}=0$ respectively. These cases would give non-trivial rules of energy orders for the multipolar modes of the entire structure.

Following with the analysis of the maps of Fig.~\ref{fig:7_HeteroD_NFs90}, the patterns also begin to hybridize low-order modes and they start including patterns corresponding to MDRs of isolated wires of higher orders when $\lambda_{eff}\rightarrow0$. The maps of \ref{fig:7_HeteroD_NFs90}(d) and (e) show complex field confinements and modulations around and inside of the high-contrast material of the wires. The spots of the intensities show multiple connections between themselves. In particular, strong field enhancements are obtained thanks to the presence of a dipolar-like mode in the biggest wire, see the panels \ref{fig:7_HeteroD_NFs90}(b) and (c).

On the other hand, the panels \ref{fig:7_HeteroD_NFs90}(b) and (c) are very related to the split peaks found in the curve in red line of \ref{fig:5_HeteroD_FF}(a). It was expressed that those peaks seemed to be split MDRs from the MDR $n_r=1$, $l_s=1$ of the wire $R=50$ nm. This mode is seen to play a key role in the maps of \ref{fig:7_HeteroD_NFs90}(b-c) as it is the main excitation in the hybridized mode.

The analysis of the near-field structures under $p$-polarization becomes more interesting. The intensities of the magnetic field reach enormous values with respect to the intensity of the illumination, see the referenced values of the saturated scales in the panels. But even more interesting is that the field enhancements can be tuned upon one wire or the other one by choosing the proper incident wavelength, see for instance (a) and (b) or (c) and (d) of Fig.~\ref{fig:8_HeteroD_NFp0}. Of course that the patterns are not symmetric, as said, for heterodimers and they go including higher-order modes for higher energies; but the ``alternacy'' of the enhancement locations on the inner fields can also be seen when comparing (a) and (b) or (c) and (d) of Fig.~\ref{fig:9_HeteroD_NFp90}. This behavior with the energy of the system will be seen to alter the dynamics of the dimer because it ``plays'' with the ``optical inertia'' induced on the wires.

Another remarkable effect occurring in the silicon dimers is the strong electric fields that can be obtained for the design of applications due to the presence of MDRs. For instance, the intensity of the electric field around the gap region is around six times the value of the incident wave, see \ref{fig:3_HomoD}(d). Similarly, strong magnetic fields can also be obtained like in \ref{fig:8_HeteroD_NFp0}(b). Although high-dielectric wires produce volume resonances that confine the fields inside the wires, the enhancements are so big that they compete with those obtained by plasmonic structures or systems that can have surface resonances \cite{Decker2016}. Even more, the coupled wires of silicon provide great intensities for both electric and magnetic fields, or for the two fundamental polarizations, which is not the usual case with 2D plasmonic structures \cite{Maier}.

\subsubsection*{Heterodimers' forces under illumination with $\varphi_0=0$ deg}

\begin{figure*}[!h]
 \begin{centering}
 \includegraphics[width=12cm,height=12cm,keepaspectratio]{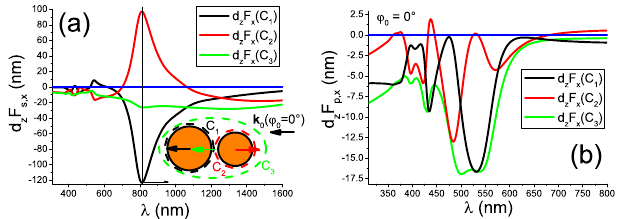} 
 \par\end{centering}
 \caption{\label{fig:10_HeteroD_Fcs0}  Optical forces exerted on the Si heterodimer of radii $r_1=50$ nm, $r_2=40$ nm and $d=5$ nm under plane-wave illumination with linear polarization. The incident angle is $\varphi_0=0$ deg. (a) and (b) Spectra of density of forces under polarization $s$ and $p$ respectively. The zeros of the scales are highlighted by blue lines.}
\end{figure*}

The mechanical response induced on the heterodimer by the configuration $\varphi_0=0$ deg is shown in the Fig.~\ref{fig:10_HeteroD_Fcs0} for the polarizations $s$ (a) and $p$ (b). As the illumination saves the symmetry of the system, the induced forces are always along the $x$-axis and of course, the induced net torques are identically zero. The illumination reaches the system by the right and the smallest wire, $2$, feel an advanced field phase with respect to that for the biggest wire, $1$. This is why the net densities of forces are different for each wire. Binding forces are overlapped with scattering forces for this configuration. However, a net scattering force is distinguished by the methodology of calculation for the whole system, see green curves in \ref{fig:10_HeteroD_Fcs0}(a) and (b). The information of the interaction given by these force components is very interesting and it complements the studies that were done in near- and far-fields. For instance, the first MDRs under $s$-polarization do not correspond to any bonding or antibonding mode, observe \ref{fig:10_HeteroD_Fcs0}(a) around $\lambda\approx1285$ nm. The corresponding map is given in panel \ref{fig:6_HeteroD_NFs0}(a). Observe that the red and black curves in \ref{fig:10_HeteroD_Fcs0}(a) have a crossing point in common which means they share a common force value. Then, the wires would be accelerated in the forward direction with respect to the illumination without attraction/repulsion between them. Simultaneously, the system moves also in the forward direction by radiation pressure; the minimum of the green curve occur for $\lambda=1285$ nm. 

The next mode, well ``detected'' by the forces, corresponds to the strong MDR occurring at $\lambda=790$ nm, see \ref{fig:6_HeteroD_NFs0}(b) and \ref{fig:5_HeteroD_FF}(a). The sign of the curves at this wavelength indicates that the MDR give place to an antibonding mode, as shown by the inset scheme in \ref{fig:10_HeteroD_Fcs0}(a). Thus the system moves as pushed by radiation pressure while each wire is simultaneously repelled from the other one at this energy. The result seems quite natural when the map (b) of the Fig.~\ref{fig:6_HeteroD_NFs0} is observed since the pattern resembles the electronic distribution of an antibonding molecular mode in diatomic molecules \cite{Demtroder}. 

The next MDR for this configuration, located at $\lambda=555$ nm in the curves of \ref{fig:10_HeteroD_Fcs0}(a), is also related with its corresponding near-field map, \ref{fig:6_HeteroD_NFs0}(c). This map shows a bonding mode; it also resembles the electronic distribution of a bonding molecular mode \cite{Demtroder}. A hot spot of the field distribution is reached in the gap region. The signs of the curves of forces, black line and red line in the online version, indicate the bonding nature of this MDR in the spectra of \ref{fig:10_HeteroD_Fcs0}(a). 

The next MDRs enter in the spectra of induced forces as the energy grows but with decreasing absolute values. Attractive or repulsive modes enter in the spectra as the sign of the coordinated forces changes. This is a characteristic of the mechanical observables; they represent more information of the resonances than the peaks in far-field curves because the sign of the forces provides information of the modes.

The forces' results for $p$-polarization bring about more curious effects. Notice that the first two MDRs, located at low energies, present an unusual force behavior in \ref{fig:10_HeteroD_Fcs0}(b). If one considers some shifts between the spectral locations of the MDRs of the far-fields and of the forces, one of the wires suffers almost zero force when the other one suffers maximum value of exerted force at resonance. For the first MDR, the stopped wire is the labeled as $2$ while for the second MDR, the stopped wire is the labeled as $1$. This manifestation is coherent with the first two ``alternating'' modes of Fig.~\ref{fig:8_HeteroD_NFp0}, i.e. panels (a) and (b) respectively. Although there exist some shifts in the resonant locations of these modes, the relation between near-field maps, forces and far-field curves is coherent. Observe that the net force is around zero on one wire when the inner-field concentration is minimum on this wire, see \ref{fig:8_HeteroD_NFp0}(a-b). The maximum field concentration for these MDRs produces the affected wire to be accelerated although the coupled neighbor is almost stopped. Of course, the force for the whole system is almost the force exerted on the wire with the focused field. Thus, the force effect of the first MDR under $\varphi_0=0$ deg is like if the wire $1$ were pulling the wire $2$, \ref{fig:8_HeteroD_NFp0}(a), while in the situation of \ref{fig:8_HeteroD_NFp0}(b), it seems like the wire $2$ would have to push the wire $1$.

When the energy is increased around $\lambda\simeq430-438$ nm, the following MDR produces a resonant positive peak in the red curve of \ref{fig:10_HeteroD_Fcs0}(b) and a negative peak or resonant dip in the black curve. A repulsion state appears, namely, an antibonding mode. This repulsion corresponds to the situation of the map \ref{fig:8_HeteroD_NFp0}(c). The following MDR is located at $\lambda=418-422$ nm in the spectra of forces and it has no direct relation with any map of the Fig.~\ref{fig:8_HeteroD_NFp0}. This results quite natural since more resolution of peaks is expected for forces' spectra than spectra obtained by far-fields. The overlapping that occurs in far-field curves for this energy region may hide this excitation. The next MDR occurs in the forces at $\lambda=395$ nm, see \ref{fig:10_HeteroD_Fcs0}(b), and it has direct correspondence with the map of \ref{fig:8_HeteroD_NFp0}(d). The difference in the absolute value of the excitations at $\lambda\simeq430-438$ nm and $\lambda=418-422$ nm indicate ``relative'' attraction states between the wires, \ref{fig:10_HeteroD_Fcs0}(b). The next resonance in \ref{fig:10_HeteroD_Fcs0}(b) is a higher-energy mode which also plays a role in the mechanical magnitudes.

\subsubsection*{Heterodimers' forces under illumination with $\varphi_0=90$ deg and $s$-polarization}

\begin{figure*}[!h]
 \begin{centering}
 \includegraphics[width=12cm,height=12cm,keepaspectratio]{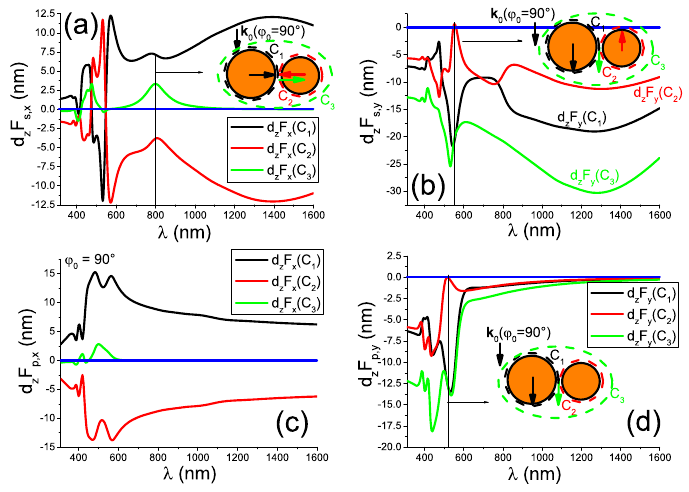}
 \par\end{centering}
 \caption{\label{fig:11_HeteroD_Fcs90}  Optical forces exerted on the Si heterodimer of radii $r_1=50$ nm, $r_2=40$ nm and $d=5$ nm under plane-wave illumination with linear polarization. The incident angle is $\varphi_0=90$ deg. (a) and (b) Spectra of $x$-components and $y$-components of density of forces under $s$-polarization, respectively. (c) and (d)  idem (a) and (b) for $p$-polarization. The zeros of the scales are highlighted by blue lines.}
\end{figure*}

The results of the induced forces under configuration $\varphi_0=90$ deg are provided in the Fig.~\ref{fig:11_HeteroD_Fcs90}. The panels \ref{fig:11_HeteroD_Fcs90}(a-b) show the binding forces and the scattering components for $s$-polarization, respectively. The panels \ref{fig:11_HeteroD_Fcs90}(c-d) show the same for $p$-polarization. The configuration $\varphi_0=90$ deg is the most relevant of the present study as more induced torques appear in the symmetry-broken system than in homodimers, see Fig.~\ref{fig:12_HeteroD_Tqs90}. Notice that all the induced peaks of forces are of comparable order under $s$-polarization, differently from the inductions seen in \ref{fig:10_HeteroD_Fcs0}(a) where a peak is relatively much stronger than the rest. In general, the induced forces under configuration $\varphi_0=90$ deg show more unusual properties than the previous configurations where the MDRs are excited, see the resonant spectral locations in Fig.~\ref{fig:11_HeteroD_Fcs90}. In particular, the validity of the action-reaction law can be evaluated for the binding forces under this symmetric illumination, as it was analyzed for plasmonic systems in \cite{Abraham2018Ag}.

The first MDR under $s$-polarization is ``felt'' by the binding forces at around $\lambda=1390$ nm, compare against the far-field resonant location that is $\lambda=1280$ nm, see \ref{fig:5_HeteroD_FF}(a). Red-shifts are expected for near-field calculations with respect to the far-field MDRs \cite{Gutierrez2016,Yuffa2016}. For the first MDR, the curves in black line and in red line are almost equal but opposite, giving zero contribution to the $x$-force for the system. In other words, the bound system follows action-reaction for this energy while it is being pushed down by radiation pressure, see \ref{fig:11_HeteroD_Fcs90}(b) at this wavelength. From the point of view of binding forces, the first mode of the system preserves the symmetry although the natural symmetry is broken, see \ref{fig:7_HeteroD_NFs90}(a). In other words, the system does not feel any lateral force for this mode.

The second mode that appears in the spectra of \ref{fig:11_HeteroD_Fcs90}(a) is given around $\lambda=780-800$ nm, depending on which curve is analyzed. If compared against the previous results, one may conclude that this mode corresponds only to the configuration $\varphi_0=0$ deg, see panels \ref{fig:6_HeteroD_NFs0}(b) and \ref{fig:10_HeteroD_Fcs0}(a). As found in previous works for plasmonic systems, the forces have enough resolution to ``detect'' peaks of excitations due to configuration $\varphi_0=0$ deg under an illumination with $\varphi_0=90$ deg. This phenomenon is a consequence of the evanescent waves present in the multiple scattering between the wires. The evanescent waves save information of the interaction regardless of the specific conditions of the illumination \cite{Girard2000}. Furthermore, under this excitation there exists a breaking in the action-reaction principle which gives place to a net lateral force for the whole system, see the green curve in \ref{fig:11_HeteroD_Fcs90}(a) around $\lambda=800$ nm. The general situation of the resultant binding forces is given by the inset scheme in \ref{fig:11_HeteroD_Fcs90}(a) for this particular resonance. On the other hand, the radiation pressure or scattering force may not appear as resonant for this wavelength, see \ref{fig:11_HeteroD_Fcs90}(b). This is probably due to that the overlapping effects mask the corresponding excitation.

From the panel \ref{fig:11_HeteroD_Fcs90}(a), a third MDR can be seen to appear at $\lambda=572$ nm. In the far field, this MDR is seen at $\lambda=565$ nm, \ref{fig:5_HeteroD_FF}(a). For this mode, there is no resultant $x$-force for the whole dimer (green curve in \ref{fig:11_HeteroD_Fcs90}(a)). This mode preserves the action-reaction law. The resonance corresponds to the pattern shown in \ref{fig:7_HeteroD_NFs90}(b) in order to establish a relation with the near field. However, the relation of this MDR with its corresponding scattering force results difficult, because the closest realistic excitation appears in \ref{fig:11_HeteroD_Fcs90}(b) at $\lambda=530-555$ nm depending on which curve is observed ($\lambda=530$ nm for the green curve). That is, the fourth and fifth MDRs can be distinguished to appear at $\lambda=532-538$ nm and $\lambda=472-484$ nm respectively in \ref{fig:11_HeteroD_Fcs90}(a). Following the maps of near-fields, the closest resonances that were obtained from far fields occur at $\lambda=525$ nm and $\lambda=475$ nm, panels \ref{fig:7_HeteroD_NFs90}(c) and (d) respectively. The overlapping and the shifts of the excitations make the desired relation difficult to apply between the black, red and green curves of the two panels (a-b) of Fig.~\ref{fig:11_HeteroD_Fcs90}. However, a very interesting result is obtained in \ref{fig:11_HeteroD_Fcs90}(b) for the MDR around $\lambda=555$ nm. An almost vanishing pulling force is obtained on the wire $2$ under plane-wave illumination and $s$-polarization, see inset scheme on the panel for graphical clarification of the effect. Logically, this behavior will lead us to relatively strong orbital torques, see below in Fig.~\ref{fig:12_HeteroD_Tqs90} and its subsequent analysis.

The bonding property of the first three MDRs can also be noticed by the sign of the curves at resonances in \ref{fig:11_HeteroD_Fcs90}(a). The fourth distinguishable MDR in \ref{fig:11_HeteroD_Fcs90}(a) changes the signs of the curves, giving an antibonding mode at $\lambda=530$ nm. The abrupt transition between the third and the fourth MDRs in \ref{fig:11_HeteroD_Fcs90}(a) can be understood by comparing the maps from the panels (b) and (c) of Fig.~\ref{fig:7_HeteroD_NFs90}. In the former map, \ref{fig:7_HeteroD_NFs90}(b), the system seems bound by the field structure while, in the latter map, the antibonding ligation can be linked to the appearance of the isolated spot inside the wire $2$. Notice the absence of the field around this spot in contrast to the field penetration inside the same wire on \ref{fig:7_HeteroD_NFs90}(b). In addition, at this fourth resonance, the system has also a net lateral force which results negative, see green curve at $\lambda=538$ nm. Thus, another breaking of action-reaction is made and the dimer would be now accelerated back into $-x$ direction. 

At higher energies, the binding forces include more MDRs in \ref{fig:11_HeteroD_Fcs90}(a) and (b) before the highest-energy excitation that is visible between $\lambda=370-385$ nm. The relation between the curves results complex again because of the overlapping peaks and the shifts between the excitations of the different curves. However, another remarkable effect is the successive excitations that are manifested in the green curve for the whole system. Each MDR that is excited originates a breaking in the action-reaction law and pushes the dimer along $x$ or $-x$ direction while it is being pushed also by radiation pressure along $-y$ direction. This is an interesting phenomenon because it would allow distinguishing homodimers from heterodimers in a hypothetical experiment with mixed systems. The deviation from the forward trajectories with respect to the incident direction $\varphi_0=90$ deg would be a signal of the present heterodimers, see Fig.~\ref{fig:4_HomoD_Fcs} for a comparison with homodimer's induced forces.

\subsubsection*{Heterodimers' forces under illumination with $\varphi_0=90$ deg and $p$-polarization}

The curves of the forces may result easier to comprehend under $p$-polarization than the curves for $s$-polarization, see panels (c) and (d) of Fig.~\ref{fig:11_HeteroD_Fcs90}. There are less overlapping of the MDRs. Furthermore, note in \ref{fig:11_HeteroD_Fcs90}(c) that the MDRs excited give only bonding modes, the black and red curves do not change signs in all the spectrum of energies. The phenomenon occurs also in the example shown for homodimers, see \ref{fig:4_HomoD_Fcs}(b). 

Similarly to the results found under $s$-polarization, the first MDR do not induce resonant $x$-forces for the system (green curve at $\lambda=555-565$ nm) but the following MDRs at increasing energies do. The appearance of higher-energy modes when the energy grows induce changes in the sign of the green curve which means lateral acceleration to the left or right for the whole dimer. This also means unbalanced forces for each wire and the consequent breaking of the Newtons' third law. At the same time, variations occur in the resonant radiation pressure for the system and for each wire, see curves of \ref{fig:11_HeteroD_Fcs90}(d). In particular, observe that a first MDR is induced at around $\lambda=535$ nm in the radiation pressure for the system (green curve in panel \ref{fig:11_HeteroD_Fcs90}(d)). Notably, a zero value of $y$-force is induced for the wire $2$ at the location $\lambda=520$ nm, see the inset scheme included for clarification. The maps corresponding to the closest resonant wavelengths are those from the first MDRs in panels (a) of Figs.~\ref{fig:8_HeteroD_NFp0} and \ref{fig:9_HeteroD_NFp90}. The other maps correspond to resonances located at wavelengths below $500$ nm. Thus, if the forces' results are related with the maps, the first MDR in \ref{fig:9_HeteroD_NFp90}(a) seems to induce relatively strong differences in the net forces for each wire and strong orbital torques (see curves of \ref{fig:12_HeteroD_Tqs90}(b) and their subsequent analysis). This would be a result of the hot spot induced on the wire $1$ of the dimer. Furthermore, it is a consequence of the geometric difference between the wires. The wire $2$ is too small to include a MDR inside itself, or the system allows for the existence of a mode inside the wire $1$ only. 

Again, the relation between the curves of panels (c) and (d) of Fig.~\ref{fig:11_HeteroD_Fcs90} is difficult to deal with at higher energies. The shifts in the resonant peaks of the different curves make the MDRs do not match themselves or between the near-field maps. However, the appearance of the MDRs in the binding and scattering forces is easy to see in the spectra. As a conclusion, a correct design of the dynamical properties of the dimer could be made from the knowledge of the geometrical and constitutive construction of it.

\subsubsection*{Heterodimers' torques}

\begin{figure*}[!h]
 \begin{centering}
 \includegraphics[width=15cm,height=15cm,keepaspectratio]{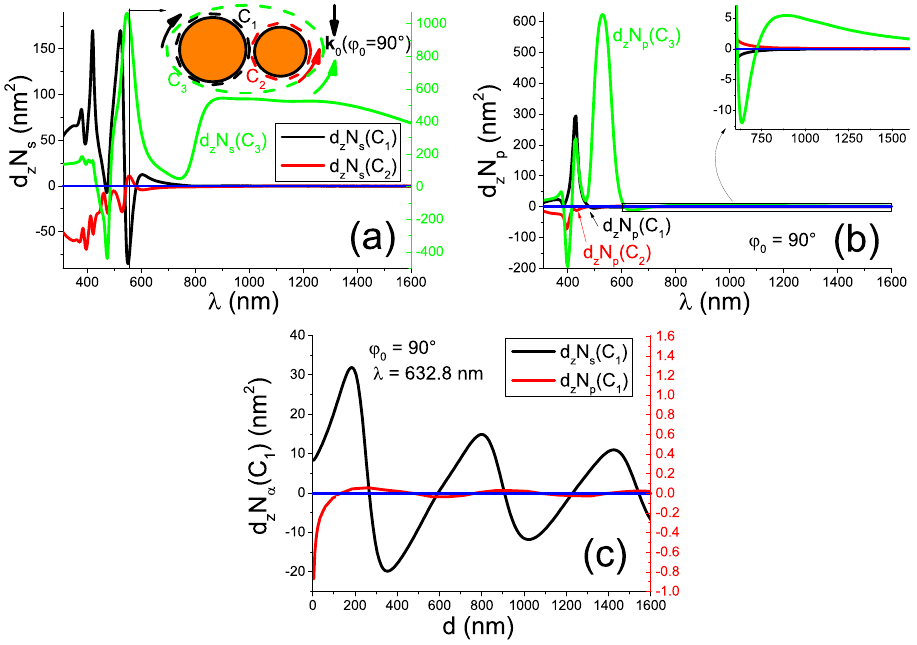}
 \par\end{centering}
 \caption{\label{fig:12_HeteroD_Tqs90}  Optical torques induced on the Si heterodimer of radii $r_1=50$ nm, $r_2=40$ nm under plane-wave illumination with linear polarization. The incident angle is $\varphi_0=90$ deg. (a) and (b) Spectra of density of torques under polarization $s$ and $p$ respectively, and $d=5$ nm. (c) Spin torque exerted on the wire 1 when varying the gap $d$. The zeros of the scales are highlighted by blue lines.}
\end{figure*}

As anticipated, the induced torques correspond to the previous results of forces and fields of the heterodimer, Fig.~\ref{fig:12_HeteroD_Tqs90}. Panel \ref{fig:12_HeteroD_Tqs90}(a) and (b) show the results under polarization $s$ and $p$ respectively. Here, the curves in black line and red line represent spin torques induced on wires $1$ and $2$ respectively and the green curves, which are not identically zero, represent orbital torques for the entire dimer. A particular situation at $\lambda=550$ nm is represented with the inset scheme on \ref{fig:12_HeteroD_Tqs90}(a). The green scale at right in \ref{fig:12_HeteroD_Tqs90}(a) corresponds to the values obtained for the green curve which reach approximatively one order de magnitude more than the values for spin torques (left ordinate scale). As a complement, the panel \ref{fig:12_HeteroD_Tqs90}(c) illustrates the behavior of the unusual spin torque induced on wire $1$ as a function of the gap between the wires. These torques are induced at the He-Ne laser wavelength of $\lambda=632.8$ nm and two curves are shown for each fundamental polarization respectively. The vertical scale on the left (right), in black (red) color, corresponds to the induced torques under $s$- ($p$-) polarization.

First, notice in \ref{fig:12_HeteroD_Tqs90}(a) that the ``monopolar'' MDR is now allowed to appear in the spectra of torques. Compare the situation against the similar homodimer configuration in \ref{fig:4_HomoD_Fcs}(c) where only spin torques exist. Furthermore, the structure of the green curve in \ref{fig:12_HeteroD_Tqs90}(a) seems to indicate that there is some overlapping of excitations at low energies (long wavelengths). The green curve under $s$-polarization, see \ref{fig:12_HeteroD_Tqs90}(a), appears to have a coupling of the peaks at $\lambda\approx1280$ nm and $\lambda\approx780-890$ nm, this latter one as a consequence of the interaction seen in far-fields at $\varphi_0=0$ deg. However, these peaks appear as resonant spin torques but with vanishing values in the black and red curves of \ref{fig:12_HeteroD_Tqs90}(a) (zoom in detail not shown here).

Due to the high resolution that offers the torques as near-field magnitudes, several narrow resonances enter in the scene for both polarizations as the energies grow, see \ref{fig:12_HeteroD_Tqs90}(a) and (b). Although these peaks are relatively narrow and very resolved, they appear as one after other. Taking also into account the shifts between the peaks from the different curves of integration, the relations between the MDRs of these curves are difficult to deal with. However, several MDRs can be recognized because they are supposed to appear in the far and near-fields at close spectral locations between them. The remarks to be made in the analysis of \ref{fig:12_HeteroD_Tqs90}(a) are that: 1- Strong orbital and spin torques are induced at the interval $\lambda\in(522-565)$ nm and at $\lambda\simeq474;420$ nm for $s$-polarization. 2- These inductions correspond to the resonances with strong asymmetric fields that show the maps \ref{fig:7_HeteroD_NFs90}(b) and (c) in the former case. 3- The MDR found at $\lambda\simeq474$ nm is related with the map \ref{fig:7_HeteroD_NFs90}(d). 

Under $p$-polarization, \ref{fig:12_HeteroD_Tqs90}(b), the MDRs in torques are clearly identifiable at $\lambda=635,530,430,400$ nm. The MDRs located at $\lambda=430;400$ nm are related with the field distributions of \ref{fig:9_HeteroD_NFp90}(c-d). The strongest induction occurs at $\lambda=530$ nm (big peak in the green curve for orbital torque) and it would correspond to the MDR in \ref{fig:9_HeteroD_NFp90}(a) with a strong asymmetric distribution of field. The correspondence of this orbital torque with the effect of the pulling force shown in \ref{fig:11_HeteroD_Fcs90}(b) is obtained exactly at the same wavelength.

Finally, a variation of the spin torque with the gap will be discussed as another illustration of the new optical effects presented here, \ref{fig:12_HeteroD_Tqs90}(c). The range of gaps goes from the studied value $d=5$ nm up to $d=1600$ nm. Notice that the laser wavelength is very near to the minimum found for the orbital torque (green curve) at $\lambda=635$ nm under $p$-polarization in \ref{fig:12_HeteroD_Tqs90}(b), see in particular the inset graphic. This results in an absolute minimum for the $p$-spin torque at $d=5$ nm, see curve in red line in \ref{fig:12_HeteroD_Tqs90}(c). Surprisingly, the spin presents damped oscillating behavior with the gap around the zero value (blue line) for both polarizations. In particular, the maximum spin is not reached at $d=5$ nm under illumination with $s$-polarization, see curve in black line. The absolute maximum reached under $s$-polarization occurs at $d=185$ nm for this example. Furthermore, the black curve has several extremals and several zeros (compare against the blue line) at specific values. Both curves present zero torque when $d\rightarrow\infty$ as the physical limit of isolated wires is reached. On the other hand, the $s$- and $p$- curves show a different behavior at the near-field distances.

There is an interesting difference when comparing the response of the spin torques due to high-dielectric vs metallic dimers. When having metallic dimers, the spin torques decay rapidly in absolute value with the increasing gap (not shown here). The response is quite different in the example shown for silicon heterodimers as the spin holds for gap distances equivalent to many characteristic wavelengths under $s$-polarization. This essential difference in the mechanical responses is due to the nature of the resonances excited in each case. For the case of metallic systems, the resonances obtained correspond to surface modes. For the case of dimers made with high-dielectrics, the resonances excited correspond to volume modes. The field concentration for the latter modes lies mainly inside the particles. However, the near fields spread out for $s$-polarization. Thus, the spin torques can be sustained along great distances. Nonetheless, the configuration $\varphi_0=90$ deg is unstable in all the spectra of dimers. The system will try to align itself with the illumination direction. In particular for homodimers, the contributions of orbital torques arise as soon as the configuration $\varphi_0=90$ deg is left. For heterodimers, the contribution of the orbital torques always exists for incident angles $\varphi_0\neq0;180$ deg so the system is even more unstable.

\section*{Conclusions}

In this work, new mechanical effects have been explored on coupled wires made of high-dielectric. In particular, the results were illustrated on two-dimensional silicon dimers under plane-wave illumination with linear polarization. The unusual inductions can only be seen with realistic calculations that include multiple scattering. Some of the effects of including multipolar radiation may involve pulling forces, broken action-reaction law, lateral forces for the entire dimer or spin and orbital torques in addition to the well-known binding and scattering forces. 

The mechanical inductions have been related with the morphology-dependent resonances (MDRs) that can be excited in systems with a high-dielectric material. The study has included the response by single wires as a way to introduce the  MDRs and their complex mechanical inductions when the wires are coupled. In particular, strong anisotropic fields are obtained in the heterodimer systems due to the excitation of MDRs that induce strong torque components in the structure.

The spin torques at illumination configurations with $\varphi_0\neq0;180$ deg are unstable because orbital torques also exist for the system (only the cases $\varphi_0=0;90$ deg have been shown here). These orbital torques would make the system to rotate towards an alignment with the illumination. As a consequence, the system may have oscillating rotation around the illumination direction while it is being pushed by radiation's pressure; and it also may be accelerating to lateral directions if it has dissimilar particles like in the heterodimer illustrated here. In the particular case of alignment with the illumination, i.e. $\varphi_0=0;180$ deg, all the torques cease. 

In particular, the unusual spin torques observed in the coupled wires constitute a new approach to the movement of the system and should be taken into account for the design of photonic-based nanodevices as, for instance, filters of nanoparticle's systems or a "nanofactory" \cite{Raziman2015}. The exhaustive study presented in this paper closes previous studies where similar effects had been found in plasmonic systems \cite{Abraham2015_1,Abraham2015_2,Abraham2016,Abraham2018Ag}. 

\begin{acknowledgments} 
The author would like to thank Marcelo Lester for sharing interesting discussions on the topic.
\end{acknowledgments}


\end{document}